\documentclass[twocolumn,english,prl,aps,superscriptaddress]{revtex4-2}
\usepackage[T1]{fontenc}%\usepackage[utf8]{inputenc}
\usepackage{color}
\usepackage{babel}
\usepackage{mathtools}
\usepackage{bm}
\usepackage{graphicx}
\usepackage{esint}
\usepackage{amsfonts}
\usepackage{graphicx}
\usepackage{float}
\usepackage{subfigure}
\usepackage{amsthm}
\usepackage{amssymb}
\usepackage{braket}
\usepackage[usenames,divipsnames,svgnames,table]{xcolor}
\usepackage[unicode=true,pdfusetitle,
bookmarks=true,bookmarksnumbered=false,bookmarksopen=false,
breaklinks=true,pdfborder={0 0 0},backref=false,colorlinks=true]{hyperref}
\hypersetup{linkcolor=NavyBlue,urlcolor=NavyBlue,citecolor=NavyBlue}
\usepackage{breakurl}

\begin{document}

\title {Quantum multiparameter estimation with graph states}

\author{Hong Tao}
\email{htaophys@hust.edu.cn}
\affiliation{College of Information Science and Technology, Jinan University, Guangzhou 510632, China}
\author{ Xiaoqing Tan}
\email{ttanxq@jnu.edu.cn}
\affiliation{College of Information Science and Technology, Jinan University, Guangzhou 510632, China}
%\date{\today}

\begin{abstract}
  In the $\mathrm{SU}(2)$ dynamics, it is especially significant to achieve a simultaneous optimal multiparameter estimation but it is very difficult. Evolution on $\mathrm{SU}(N)$ dynamics is a research method to explore simultaneous multiparameter estimation with the quantum network.
  As the highly entangled states, graph state, is an intrinsical quantum resource for quantum metrology. For $n$-qubit graph state, we propose a simultaneous multiparameter estimation scheme that investigates evolution in $\mathrm{SU}(N)$ dynamics.
  For single-parameter estimation, the precision limit beyond the Heisenberg limit in the higher dimension spin of $\mathrm{SU}(2)$.
  We consider two scenarios where the Hamiltonian operator is commutation and non-commutation respectively and verify that the global estimation precision is higher than the local estimation precision. In the parameter limit condition,  the precision of parameter estimation for the simultaneous estimation of each parameter is equal to the precision of the single-parameter estimation. In addition, we find a precision-enhancement scheme that depends on the dynamics $\mathrm{SU}(N)$.
  The smaller the $N$ for the dynamics evolution, the higher the precision of the parameter estimation. Finally, we prove that the graph state is the optimal state of quantum metrology, a set of optimal measurement basic can be found, and the precision limit of multiparameter estimation can attain the quantum Cram\'{e}r-Rao bound.

\end{abstract}

\maketitle

\section{introduction}
In many-body quantum systems, the interaction between physical systems can be directly identified for many cases, such as the next-neighbor interactions in a coupled system. Such a quantum system can generally be represented by a quantum network $G=(V, E)$, which can mathematically, it can be seen as a graph shape whose vertices $V$ correspond to physical systems and edges $E$ represent interactions~\cite{West2001,Diestel2000}.

\begin{figure}[tp]
\includegraphics[width=8.5cm]{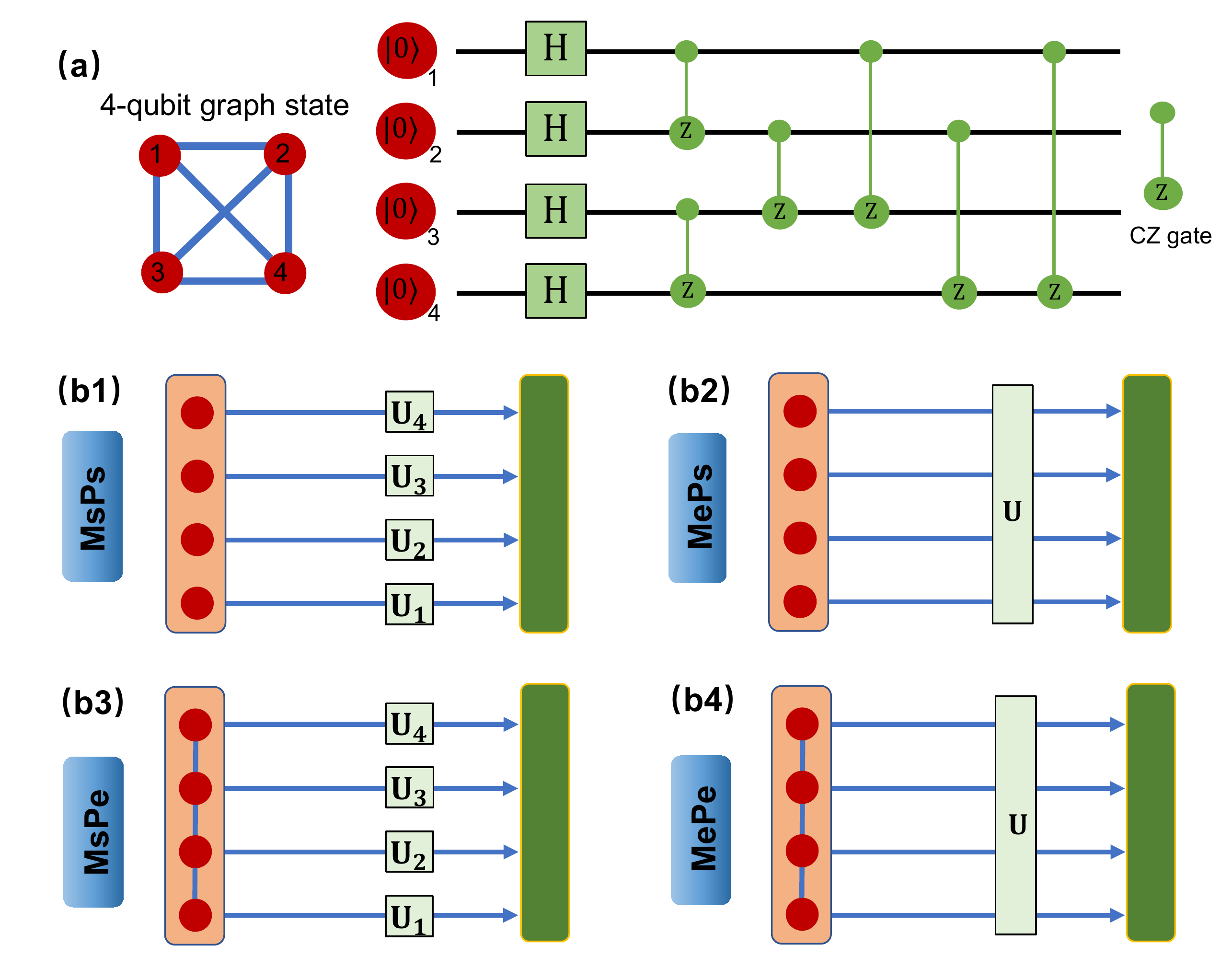} \centering
\caption{ (a) Illustrate the quantum circuit for the four-qubit complete graph states. The red vertices are the qubits of the quantum network and the blue line is the interaction between the next-neighbor qubits. The Hadamard gate and the control-Z gates are used to prepare the graph state. The Hadamard gate is $\mathrm{H}=\frac{1}{\sqrt{2}}(\sigma_z +\sigma_x)$. The CZ gate is the control-$Z$ gate
$\mathrm{CZ}\equiv\left(\openone^{\otimes 2}-|11\rangle\langle 11|\right) \otimes \openone+|11\rangle\langle 11| \otimes \sigma_z$. (b) Discretized quantum multiparameter estimation schemes (blue)  including the mode and particle separable (MsPs),
the mode separable and particle entangled (MsPe),
the mode entangled and particle separable (MePs),
the mode and particle entangled (MePe).
We consider the state preparation (red), the unitary parameterization processing (green), and the projection measurement (dark green).} \label{Fig:Est_scheme}
\end{figure}

The quantum network is regarded as the quantum state of the physical system, that is, the graph state is the quantum state of the physical system, including multiple components~\cite{Raussendorf2003,Aschauer2003,Grassl2002}. A graph can be thought of as a pattern of interaction: when two particles, such as the spin-1/2 system, interact through an interaction (such as the quantum Ising system), the graph connecting the two related vertices has an edge. We can think of the adjacency matrix of a simple graph as a symmetric $n \times n$ matrix of a system composed of $n$-qubit, whose elements are taken from $\{0,~1\}$ \cite{Raussendorf2003,Aschauer2003,Briegel2001,Grassl2002,Hein2004,Aschauer2005,NEST2004}, from the graph-theoretical view, it is an undirected graph. In this sense, the graph can be viewed as an interaction between particles~\cite{NEST2004}.
\begin{equation}
  |G\rangle=\prod_{\{a,b\} \in E}U_{a,b}|+\rangle^{V},
\end{equation}
where the quantum state $|+\rangle$ is the eigenstate of the Pauli-$X$ with eigenvalue $+1$, and the phase gate $U_{a,b}$ is applied to all vertices $a, b$ that are adjacent in graph $G$ as shown in Fig.~\ref{Fig:Est_scheme}(a) (Illustrate the quantum circuit for the four-qubit complete graph states, the red vertices are the qubits of
the quantum network and the blue line is the interaction between the next-neighbor qubits. The Hadamard gate and the
control-$Z$ gates are used to prepare the graph state. The Hadamard gate is $\mathrm{H}=\frac{1}{\sqrt{2}}(\sigma_z +\sigma_x)$, and the CZ gate is the control-$Z$ gate
$\mathrm{CZ}\equiv\left(\openone^{\otimes 2}-|11\rangle\langle 11|\right) \otimes \openone+|11\rangle\langle 11| \otimes \sigma_z$).  At the same time, the adjacency matrix encodes the stabilizer $g$ of the quantum state, that is, the complete set of eigenvalue equations satisfied by the quantum state. Therefore, graph states are actually stabilizer states~\cite{Gottesman1997}. Such graphs play a central role in quantum information theory.

The quantum network used in this paper is the special pure quantum state of the distributed quantum system. It corresponds to the graph structure, where each edge represents the Ising interaction between pairs of quantum spin systems or qubits (assuming that the element of the weight matrix of the graph is the same, with a weight value of one). Special examples of graph states are various quantum error correction codes~\cite{Schlingemann2002}, such graphs have topological protection structures that are crucial in protecting quantum states from decoherence in quantum computing~\cite{Gottesman1997}. Other examples are Greenberger-Horne-Zeilinger (GHZ) states and cluster states of arbitrary dimensions, applied in quantum communication, and quantum metrology, which are known to serve as general resources for quantum computing in the one-way quantum computers~\cite{Briegel2001,Raussendorf2001} and as intrinsical resources for quantum metrology~\cite{Shettell2020}.

Quantum metrology describes quantum parameter estimation beyond the precision of classical parameter estimation~\cite{Liu2020,Szczykulska2016,Liu2022,Albare2020,Apellaniz2016,Giovannetti2006,Giovannetti2004,Giovannetti2011,Demkowicz2015,Demkowicz2012}.
classical parameter estimation can estimate the unknown parameter $\theta$ by using $n$-qubit quantum separable states through this parameter estimation scheme. The highest precision of parameter estimation can be obtained, described as the mean squared error $\delta^2\theta$ is inversely proportional to the number of quantum states $n$, $\delta^2\theta \geq 1/n$, also known as the standard quantum limit (SQL). However, with $n$-qubit quantum entangled states, the highest precision of quantum parameter estimation can attain  $\delta^2\theta \geq 1/n^2$, also known as the Heisenberg limit (HL).

Quantum resources provide a framework beyond the precision of classical parameter estimation, and quantum networks, as important quantum resources with quantum metrology, can be used to describe most quantum metrology problems, including mapping magnetic fields~\cite{Pham2011,Steinert2010,Hall2012,Seo2007,Baumgratz2016}, phase imaging~\cite{Liu2016,Humphreys2013,Knott2016,Zhang2017,Gagatsos2016,Yue2014,Ciampini2016}, and global frequency standards~\cite{Komar2014}, etc., have conducted in-depth research both theoretically and experimentally. Although most of these studies focus on the single-phase parameter estimation, applications generally involve estimating multiple parameters simultaneously. While the quantum precision limit of single-phase parameter estimation is always achievable by the quantum Cram\'{e}r-Rao bound, the quantum precision limits of multiparameter phase estimation are not necessarily achievable~\cite{Proctor2018,Gessner2018,Yuan2016,Imai2007,Pezz2017,Goldberg2021,Triggiani2019,Rubio2020,Albarelli2020,Chen2022,Gross2021}. This makes multiparameter phase estimation necessary in the study of quantum metrology.

A major theoretical tool in quantum multiparameter estimation is the quantum
Cram\'{e}r-Rao bound (Helstrom bound)~\cite{Helstrom1976,Holevo1982}, in which
the covariance matrix of a vector of unknown parameters $\bm{\theta}=(\theta_1,\theta_2,\cdots,\theta_d)$
is lower bounded by the quantum Fisher information matrix (QFIM) $\mathcal{F}$. The entry of
QFIM is defined by $\mathcal{F}_{ij}=\mathrm{Tr}(\rho \{L_i,L_j\})/2$, where
$\{,\cdot,\}$ is the anti-commutation and $L_i$ is determined by $\partial_{\theta_i}\rho
=(\rho L_i+L_i\rho)/2$, and called the symmetric logarithmic derivative (SLD)~\cite{Matsumoto2002}, where the partial derivative is $\partial_\theta=\partial/\partial \theta$. It can be seen that the quantum
Cram\'{e}r-Rao bound is a matrix inequality, and in general, the ultimate bound is not saturable for all parameters. This is because the corresponding optimal measurements might be incompatible~\cite{Ragy2016,Heinosaari2016,Zhu2015}, and the precision of the optimal parameter estimates for each individual parameter cannot be achieved simultaneously.

In this work, we demonstrate the precision limit for quantum multiparameter estimation in the quantum network state $\rho_0=|G\rangle\langle G|$ (see Fig.~\ref{Fig:Est_scheme}(a), the four-qubit graph state),  the precision limits are given by the quantum
Cram\'{e}r-Rao bound. Although the particle-separable strategy using mode-entanglement (MePs in Fig.~\ref{Fig:Est_scheme}(b)) improves the precision of quantum parameter estimation of the mode-separable (MsPs) of particle-separable states, mode-entanglement is very important for improving the precision of the multiparameter estimation~\cite{Gessner2018}. But mode-entanglement is not necessary to improve the precision of multiparameter estimation. In the presence of particle entanglement (MsPe), the high-precision limit can be obtained through the mode-separable state. Finally, the multiparameter Heisenberg limit can only be reached when both particle entanglement and mode-entanglement (MePe) are present.

The bounds we are discussing are saturated with quantum network states $\rho_0$.
The density matrix of the quantum network stat $\rho_0$ evolves into $\rho_{\bm{\theta}}=U_{\bm{\theta}}\rho_0 U_{\bm{\theta}}^\dagger$ under the parameter-dependent dynamics $U_{\bm{\theta}} \in \mathrm{SU}(N)$ in quantum multiparameter estimation. The ultimate estimation precision of each parameter $\theta_j$ is measured by the inverse of the quantum Fisher information (QFI). The overall precision limit of all quantum parameters estimation $\bm{\theta}$ are measured by the quantum
Cram\'{e}r-Rao bound theory, that is the inverse of the QFIM.  We give the expressions of QFIM for two modes of the MsPe and MePe, respectively in the $\mathrm{SU}(N)$ dynamics. For these two modes with $n$-qubit graph states, we now next make a detailed exploration in the following.

%======================figure================================================
%============================================================================

\section{Multiparameter estimation on SU(N)}

Consider the quantum multiparameter estimation on the graph state. The typical scenario needs $d$-dimensions vector $\bm{\theta}=(\theta_1, \theta_2, \cdots, \theta_d) \in \mathbb{R}^d$ as the parameters on the probe state $\rho_0$ ($\rho_0$ is the density matrix of the graph state which can be represented by the stabilizer.), which evolute by the unitary processing $U_{\bm{\theta}} \in \mathrm{SU}(N)$ that can encode the parameters $\bm{\theta}$. The group $\mathrm{SU}(N)$ is the special unitary Lie group. The $\mathrm{SU}(N)$ symmetry group is therefore specified by a total of $N^2-1$
standard traceless non-diagonal and diagonal symmetric and antisymmetric generators and $N-1$ non-traceless diagonal symmetric generators~\cite{Omolo2018,Haber2021}. In the fundamental representation, the generators are $N \times N$ matrices. Such as the traceless generators of $\mathrm{SU}(2)$ are the Pauli operators $
\sigma^x, \sigma^y, \sigma^z$, and the Pauli matrix of $\mathrm{SU}(2)$ is extended to $\mathrm{SU}(3)$, the Gell-Mann matrices is the traceless generators of the $\mathrm{SU}(3)$. $\mathrm{SU}(2)$ encodes spin and isospin, $\mathrm{SU}(3)$ describes both color and the physics of three-light quark flavors. Since the $\mathrm{SU}(N)$ symmetry group was a diversionary spin-off from the dynamical structure of the $N$-level atom-field interaction and characterizing generators of the $\mathrm{SU}(N)$ symmetry groups for all $N \geq 2$.  Therefore, the Pauli operator of $\mathrm{SU}(2)$ can be extended to the generators of the $\mathrm{SU}(N)$.

Now we consider a $n$-qubit graph state $G=(V, E)$, which can be defined in correspondence to a graph with $n$ vertices $V$ and edges $E$.  Examples are as shown in Fig.~\ref{Fig:Est_scheme}(a) for the four-qubit quantum graph state. Here, we assume that the graph has no isolated vertices, which means that each qubit is entangled with other qubits.
The density matrix of the graph state  can be written as
\begin{equation}
\rho_{0}=\prod_{i=1}^{n}\frac{1}{2^n}(g_{i}+\openone^{\otimes n}),
\end{equation}
where $\openone$ is a 2-dimensional identity matrix, and the stabilizer can be written as
\begin{equation}
  g_i=\sigma^x_{i} \bigotimes_{j\in\mathcal{N}(i)} \sigma^z_{j}
\end{equation}
is the stabilizer for the $i$th qubit. $\mathcal{N}(i)$ is the neighborhood of the $i$th qubit,
and the operator $\sigma^{x(z)}_i$ is the Pauli $X(Z)$ matrix for the $i$th qubit.
We consider the unitary parametrized process  with the (not necessarily commuting) operator $H_k \in \mathrm{SU}(N)$, the unitary operator can be expressed as
\begin{equation}
  U_{\bm{\theta}}=\mathrm{e}^{-i \bm{H}(\bm{\theta})}=\mathrm{e}^{-i\sum_{k=1}^d \theta  _{k } H_k },
\end{equation}
where the Hamiltonian  $\bm{H}(\bm{\theta})$ depends on the parameters $\bm{\theta}$, and $H_k$ is the generator of the $\mathrm{SU}(N)$.
Now, we request ourselves to the situation where the graph state evolves under the one qubit Hamiltonian $H_k$ for $k= 1,2,\cdots, n$. The global Hamiltonian can write as
\begin{equation}
  \bm{H}(\bm{\theta})= \sum_{k=1}^n \theta_k H_k.
\end{equation}
Now, let us consider the QFIM under the Hamiltonians and graph states discussed above. We show that QFIM can be written as the covariance matrix of the Hermitian operator $\mathcal{H}$.
\begin{align}\label{Eq:QFIM}
  \mathcal{F}_{jk}= 4 \left(\mathrm{Tr}(\mathcal{H}_j \mathcal{H}_k \rho_0)-\mathrm{Tr}(\mathcal{H}_j \rho_0)\mathrm{Tr}( \mathcal{H}_k \rho_0)\right),
\end{align}
where the Hermitian operator $\mathcal{H}_j:= i \partial_{\theta_j} U_{\bm{\theta}}^{\dagger} U_{\bm{\theta}}$ with parameter $\theta_j$,
then one can find
\begin{align}
  \mathcal{H}_j=-\sum_{m=0} \frac{i^m}{(m+1)!}(\bm{H}(\bm{\theta})^\times)^m \partial_{\theta_j}\bm{H}(\bm{\theta}),
\end{align}
where $H^\times(\bm{\cdot})= [H, ~\bm{\cdot}]$.
In general, for general unitary dynamics, the QFIM might depend on the parameters $\bm{\theta}$, and the Hermitian generators of the $\mathrm{SU}(N)$ all do not commute $[\mathcal{H}_j, \mathcal{H}_j] \neq 0$. When the generators all commute, this implies that simultaneous estimation does not provide the intrinsic advantages than individual estimation. This work can achieve the ultimate quantum precision limit. We now can state the precision limit by the quantum Cram\'{e}r-Rao bound
\begin{equation}
  \mathrm{Cov}(\bm{\theta}) \geq \frac{1}{\mu \mathcal{F}},
\end{equation}
where $\mathrm{Cov}(\bm{\theta})$ is the covariance matrix of the unknown parameters $\bm{\theta}$, $\mu$ represents the number of times that the estimation procedure is repeated. In this case, we assume the QFIM can be invertible, which means the singular QFIM can not be independent of all unknown parameters and the parameters can not be estimated simultaneously. In this paper, we take $\delta^2 \bm{\theta}:= \sum_{i=1}^{d}\delta^2\theta_i$ as the figure of the merit. The precision limit of the quantum multiparameter estimation is
\begin{equation}
  \delta^2 \bm{\theta} \geq \frac{1}{\mu}  \mathrm{Tr}(\mathcal{F}^{-1}).
\end{equation}
Improving the precision of each parameter estimation to be as close to the quantum
Cram\'{e}r-Rao bound as possible, and whether the quantum Cram\'{e}r-Rao bound can be simultaneously achieved (that is, whether multiparameter can be optimally estimated simultaneously) that is also an important issue.

\subsection{single-parameter estimation}
\begin{figure}[tp!]
\includegraphics[width=8.5cm]{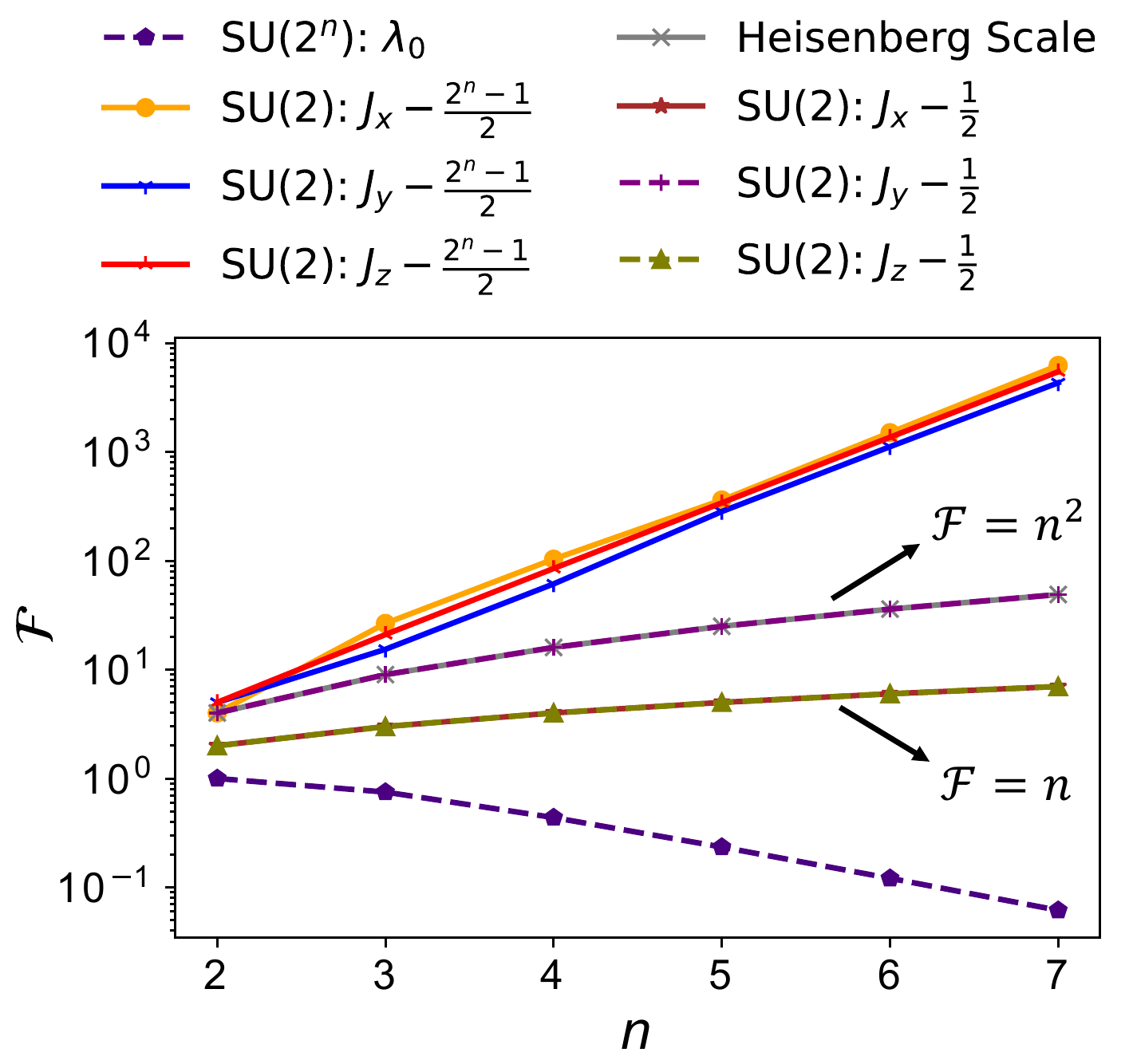} \centering
\caption{ Plot for estimation of the single parameters of the n-qubit complete graph state. The QFI as the generators $H_k$ with $\mathrm{SU}(2)$ (spin-$\frac{1}{2}$ and high-dimension spin-$\frac{2^n-1}{2}$) and SU($2^n$). The operators $J_m$, $m=\{x, y ,z\}$ are the generators of the $\mathrm{SU}(2)$ and $\lambda_0$ is the generator of the SU($2^n$). \label{Fig_signlepara}}
\end{figure}

The graph state is given in the literature~\cite{Shettell2020} as a quantum resource that can reach the Heisenberg limit for single-parameter estimates with $\mathrm{SU}(2)$.
For the unitary operator with the generator of the  $\mathrm{SU}(2)$ (The generators are the Pauli operators, $\sigma^x$, $\sigma^y$, $\sigma^z$.), the unitary operator can be written as
\begin{align}
    U_\theta=\mathrm{e}^{-i \theta H_k},
\end{align}
where $H_k=\frac{1}{2}\sum_{j=1}^n\sigma_j^k$ is the collective operator with spin-1/2, $k=\{ x, y, z\}$. The QFI is
\begin{align}
    F=\sum_{j,k=1}^{n}\mathcal{F}_{j,k}.
\end{align}
Here the $\mathcal{F}_{jk}$ is the matrix elements of the QFIM with the Appendix~\ref{Sec:individual} for the generator of the $\mathrm{SU}(2)$.
When we consider the generator $H_k$ of the high spin-$\frac{2^n-1}{2}$ with $\mathrm{SU}(2)$ and the SU($2^n$) which is the Hamiltonian is $H_k=\lambda_k$, the $\lambda_k$ is the $k$th generator of the SU($2^n$). We can find the QFI can be simplified for the graph state
\begin{align}
    \mathcal{F}=4(\mathrm{Tr}(\rho_0 H_k^2)- \mathrm{Tr}(\rho_0 H_k)^2).
\end{align}
As shown in Fig.~\ref{Fig_signlepara}, the quantum precision limit can reach the Heisenberg scaling $n^2$ when the Hamiltonian is generator $J_y$ of the  spin-1/2. When the Hamiltonian is the generator of high-dimensional spin-$\frac{2^n-1}{2}$, the accuracy of the parameter estimation is beyond the Heisenberg limit. That is, the Hamiltonian quantities of high-dimensional spins can enhance the precision of parameter estimation. However, when the Hamiltonian is the generator of the SU($2^n$), the accuracy of the parameter estimation does not reach the standard quantum limit.
%======================figure=======================
%===================================================
\subsection{Local estimation with graph state}

The different evolution modes of quantum states are shown in Fig.~\ref{Fig:Est_scheme}(b). The $n$-qubit graph state evolves through the two modes (MsPe and MePe). In this subsection, we consider the case where the evolution satisfies mode-separable with graph state.

The evolutionary dynamics satisfies unitary parameterization process $U_{\bm{\theta}}=\mathrm{e}^{-i \bm{\theta} \cdot \bm{H}}$ with the $\mathrm{SU}(N)$, where $\bm{\theta}=\{\theta_1, \theta_2, \cdots, \theta_n\}$, and
$\bm{H}=\{ H_1, H_2, \cdots, H_n \}$.
The Hermitian operator $H_j$ is the generator of the $\mathrm{SU}(N)$ ($N=2^m, ~m=\{1,2,\cdots , n\}$) and satisfies the commute with each other $[H_j, H_k]=0$ for all $j, ~k \in \{ 1,2, \cdots, n\}$. For the quantum scheme exploiting graph states where we estimate the parameters locally.

For any $n$-qubit graph state $\rho_0$ , the parameters $\theta_k$ to be estimated are the coefficients of a set of generators $H_k$ with $\mathrm{SU}(N), N \geq 2$, if $\left[H_j, H_k\right]=0$ for all $j, k$, the QFIM is  parameter independent, and the QFIM is invertible with different generators $H_j$ of the $\mathrm{SU}(N)$, then the precision limit of the parameter estimation is
$\delta^2 \bm{\theta} \geq c$, where $c=\mathrm{Tr}(\mathcal{F}^{-1})$ is a constant number.

\begin{figure}[tp!]
\includegraphics[width=8.5cm]{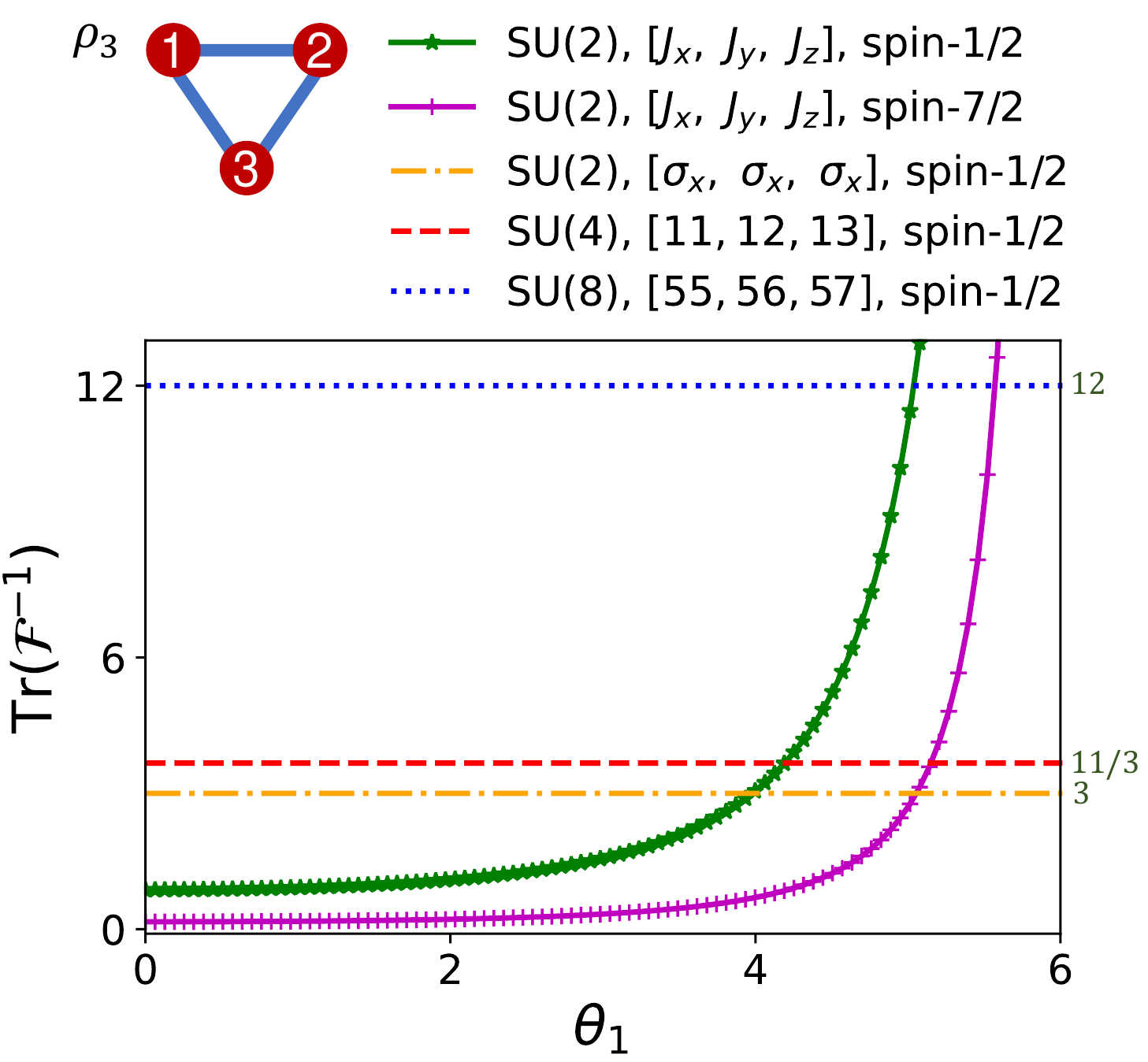} \centering
\caption{ Plot for estimation of the three parameters of the three-qubit complete graph state $\rho_3$. We show the total variance for three different dynamics scenarios with the $\mathrm{SU}(2), ~\mathrm{SU}(4),~\mathrm{SU}(8)$ for $\theta_2=\pi/3,~ \theta_3=\pi/4$. The dynamics is the $\mathrm{SU}(N)$ parameterization process $U_{\bm{\theta}}=\mathrm{e}^{-i \bm{\theta} \cdot \bm{H}}$, and the operator $H_j$ is the generator of the $\mathrm{SU}(N)$, and satisfy the commute with each other $[H_j, H_k]=0$ (blue line (SU(8)), red line (SU(4)), orange line ($\mathrm{SU}(2)$)). The precision limit of the three parameters ($\theta_1, ~\theta_2, ~\theta_3$ ) is a constant number. The $[55, ~56, ~57]$ represents the 55th (56th and 57th) generators of the SU(8) as shown in Appendix.~\ref{sec:gen}. For the SU(8), the generators  number is 63, which is $\{0, 1, \cdots, 63\}$. Similarly $[11, ~12, ~13]$ represents the generators of the SU(4), respectively. }\label{Fig_individual}
\end{figure}

A complete proof is provided in Appendix~\ref{Sec:individual}. When the operators $H_j, H_k$ are the different generators of the $\mathrm{SU}(N)$ for all $j, k \in \{1, 2, \cdots, n\}$, the QFIM is invertible. The QFIM can read
\begin{equation}
  \mathcal{F}_{jk}=4[ \mathrm{Tr}(H_j H_k \rho_0)-\mathrm{Tr}(H_j  \rho_0)\mathrm{Tr}( H_k \rho_0)].
\end{equation}
In this case, each parameter is mutually compatible, and what is the attainability of the multiparameter quantum Cram\'{e}r-Rao bound as shown in Fig.~\ref{Fig_individual}. In Fig.~\ref{Fig_individual}, we give an example that shows a three-qubit complete graph state $\rho_3$, the precision limit of the three parameters estimation is a constant number with the different dynamics scenarios $\mathrm{SU}(2), ~\mathrm{SU}(4),~\mathrm{SU}(8)$, which the Hermitian operators $H_j$ commute with each other $[H_j, H_k]=0$, the precision limits with three-parameter are $3, ~11/3, ~12$, respectively.
%======================figure================================================
%============================================================================

However, when there are operators in which $H_j$ and $H_k$ are the same generators of the $\mathrm{SU}(N)$, the QFIM is not necessarily invertible. i.e., the operators $\sigma_j^x$ and $\sigma_k^x$ are the generator $\sigma^x$ of the $\mathrm{SU}(2)$, where $\sigma_j^x$ and $\sigma_k^x$ are the Pauli $X$ on the $j$th and $k$th qubit, respectively. If the QFIM can not be full rank, which means each parameter are not mutually incompatible, that is the optimal estimate for all parameters can not be made in their common basic measurements.
Note that, the elements of the QFIM can read
\begin{equation}
  \mathcal{F}_{jk}= \mathrm{Tr} (\sigma_j^m \sigma_k^m \rho_0)
\end{equation}
for $\mathrm{SU}(2)$ in which the operator satisfies $\mathrm{Tr}(\sigma_j^m \rho_0)=0$ for $m=x, y, z$, and the QFIM is not necessarily invertible. $\sigma_j^m$ is the Pauli $m$ operators acting on the $j$th qubit. The proof shows the Appendix~\ref{sec:sigma}.
For example, we consider a $n$-parameter local estimation on $\mathrm{SU}(2)$:
 \begin{equation}\label{eq_un_sigmax}
   U_{\bm{\theta}}=\exp(-\frac{i}{2}\bm{\theta}\cdot\bm{\sigma}^x)
 \end{equation}
with the parameters $\bm{\theta}=
 (\theta_1, \theta_2,\cdots,\theta_n)$ and the Hermitian operators $\bm{\sigma}^x=(\sigma^x_1,\sigma^x_2,\cdots,\sigma^x_n)$.
 In this case, the diagonal entry of QFIM is $\mathcal{F}_{jj}=1$ for any $\theta_j$ and the off-diagonal entry reads
 \begin{equation}
 \mathcal{F}_{jk}=\begin{cases}
 1,~\mathrm{for}~\mathcal{N}(j)=\mathcal{N}(k),\\
 0,~\mathrm{for}~\mathcal{N}(j)\neq\mathcal{N}(k).
 \end{cases}
 \end{equation}
 Based on the Cram\'{e}r-Rao bound, the variance of $\theta_j$ satisfies
 $\delta^2 \hat{\theta}_j\geq[\mathcal{F}^{-1}]_{jj}\geq 1/\mathcal{F}_{jj}$ with
 $\hat{\theta}_j$ an unbiased estimator of $\theta_j$. $\mathcal{F}_{jj}$ is the
 corresponding quantum Fisher information (QFI).
 Notice the fact that $\mathcal{F}_{jj}=1$ in this case, the lowest $\delta^2 \hat{\theta}_j$ is also $1$, and can only be attained when $\mathcal{F}$ is diagonal.
 Hence, the graph states having a diagonal QFIM offer the highest precision limit.
 However, this precision limit is no better than the scheme with isolated qubits.
 For a single qubit with the parameterization process $\exp(-\frac{i}{2}\theta\sigma_x)$, the maximum QFI is also $1$ and can be obtained by the superposition of $\sigma_x$'s eigenstates.
 This fact indicates that $n$-qubit graph states do not show advantages on the $n$-parameter estimation, compared to the schemes with isolated qubits.

 Since the resulting QFIM is not necessarily invertible when the Hamiltonian operator is an operator formed by the same generator tensor of the $\mathrm{SU}(N)$, i.e., the Eq.~(\ref{eq_un_sigmax}), we give a rule whose dynamics are constrained on $\mathrm{SU}(2)$, which called the \emph{single-joint connecting rule} (SJCR) (The detailed description is in Appendix~\ref{Sec:digQFIM}.).  Under the dynamic evolution of $\mathrm{SU}(2)$, the parameterization process is described by Eq.~(\ref{eq_un_sigmax}), and this rule gives a scheme for constructing the QFIM to be an identity matrix. This scheme describes the simultaneous multiparameter estimation in the dynamic process of $\mathrm{SU}(2)$ in which the precision limit of the multiparameter estimation is $n$.

 \subsection{Global estimation with graph state}
 In this subsection, we consider the case where the evolution satisfies mode-entangled with graph state. And then, we consider the $n$-parameter global estimation on $\mathrm{SU}(N)$
 \begin{equation}
   U_{\bm{\theta}}=\exp(-i \bm{\theta}\cdot \bm{H})
 \end{equation}
 with the generator $\bm{H}= (H_1, H_2, \cdots, H_n)$, the operators $H_k$ are the generators of the $\mathrm{SU}(N)$ with $[H_j, H_k]\neq 0$. For $n$-qubit graph state,  we can select the set of parameterization process $\mathrm{SU}(N)$, such as
 \begin{equation*}
\begin{aligned}
   &2-\mathrm{qubit} : \mathrm{SU}(2), \mathrm{SU}(2^2);\\
   &3-\mathrm{qubit} : \mathrm{SU}(2), \mathrm{SU}(2^2), \mathrm{SU}(2^3);\\
   &4-\mathrm{qubit} : \mathrm{SU}(2^2), \mathrm{SU}(2^3), \mathrm{SU}(2^4);\\
   &5-\mathrm{qubit}: \mathrm{SU}(2^2), \mathrm{SU}(2^3), \mathrm{SU}(2^4), \mathrm{SU}(2^5); \\
   &6-\mathrm{qubit}: \mathrm{SU}(2^2), \mathrm{SU}(2^3), \mathrm{SU}(2^4), \mathrm{SU}(2^5), \mathrm{SU}(2^6);\\
   &\cdots \\
   &n-\mathrm{qubit}: \mathrm{SU}(N), \cdots,  \mathrm{SU}(2^n),
 \end{aligned}
 \end{equation*}
 where $N=2^m$ and $m$ is a positive integer and $m \leq n$, and when $N^2-1 \leq n$, $N$ takes the minimum value. This implies that in the parameter limit $\theta_j \rightarrow 0$, the ultimate quantum limit state by the QFIM in $\mathrm{SU}(N)$ dynamics.

 For any $n$-qubit graph state $\rho_0$, the parameters $\theta_k$ to be estimated are the coefficients of a set of generators $H_k$ for $\mathrm{SU}(2^n), n \geq 2$, if $[H_j, H_k] \neq 0$, the QFIM  is invertible and depends on the values of the parameters. In the limit $\theta_j \rightarrow 0$, the entries of QFIM can then
 be expressed by
 \begin{equation}
 \mathcal{F}_{j,k}=4[\mathrm{Tr}(H_j H_k \rho_0)- \mathrm{Tr}(H_j\rho_0)\mathrm{Tr}( H_k \rho_0)].
 \end{equation}
 The proof is provided in Appendix~\ref{Sec:simultaneous}.
 In this case, the QFIM is always invertible and depends on the estimated parameters $\bm{\theta}$ and satisfies the Eq.~(\ref{Eq:QFIM}) as shown in  Fig.~\ref{Fig_individual}.

 Here we give an example to illustrate that global estimation is more precision than the local estimation of parameters for three-qubit graph state $\rho_{3}$ as shown in Fig.~\ref{Fig_individual} with parameters on $\mathrm{SU}(2)$, the global Hamiltonian could be written as
 \begin{equation}
   \bm{H}(\bm{\theta})= \theta_1 H_1 +\theta_2 H_2 +\theta_3 H_3
 \end{equation}
 with the three parameters $\theta_{1}=$ $B \sin \theta \cos \phi,  ~ \theta_{2}=B \sin \theta \sin \phi,  ~ \theta_{3}=B \cos \theta$, where the operators $H_{1, 2, 3}=J_{x,y,z}=\frac{1}{2}\sum_{j=1}^3\sigma^{x,y,z}_j$ are the collective operators with the generators of the $\mathrm{SU}(2)$. We can find that the matrix elements of the QFIM are
 \begin{align*}
   \mathcal{F}_{11}&=4\langle J_x^2 \rangle + \theta_3^2\langle J_y^2\rangle + \theta_2^2\langle  J_z^2\rangle,\\
   \mathcal{F}_{22}&=\theta_3^2\langle J_x^2 \rangle + 4\langle J_y^2\rangle + \theta_1^2\langle  J_z^2\rangle,\\
   \mathcal{F}_{33}&=\theta_2^2\langle J_x^2 \rangle + \theta_1^2\langle J_y^2\rangle + 4\langle  J_z^2\rangle,\\
   \mathcal{F}_{12}&=2\theta_3 (\langle J_x^2 \rangle - \langle J_y^2\rangle) -\theta_1 \theta_2\langle J_z^2\rangle,\\
   \mathcal{F}_{13}&=2\theta_2 (\langle J_z^2\rangle-\langle J_x^2 \rangle )  -\theta_1 \theta_3 \langle J_y^2\rangle,\\
   \mathcal{F}_{23}&=2\theta_1 (\langle J_y^2\rangle-\langle J_z^2 \rangle )  -\theta_2 \theta_3 \langle J_x^2\rangle.
 \end{align*}
 In the parameters limit $\theta_k \rightarrow 0$, for all $k=1,2,3$, the QFIM can be rewritten into
 \begin{equation}
 \mathcal{F}=\left(\begin{array}{ccc}
 4\langle J_x^2\rangle & 0 & 0 \\
 0 & 4\langle J_y^2\rangle & 0 \\
 0 & 0 & 4\langle J_z^2\rangle
 \end{array}\right).
 \end{equation}
 The highest precision of the three-qubit complete graph state satisfies $\delta^2 \bm{\theta}\rightarrow \frac{2n+1}{n^2}|_{n=3}$ with the three parameters $\bm{\theta}=\{ \theta_1, \theta_2, \theta_3\}$.
 However, we consider the local Hamiltonian
 \begin{equation}
   \bm{H}(\bm{\theta})= \frac{1}{2}(\theta_1 \sigma^x_1 +\theta_2  \sigma^x_2 +\theta_3  \sigma^x_3).
 \end{equation}
 The precision limit of the three-qubit complete graph state is $\delta^2 \bm{\theta}=n|_{n=3}$ in Fig.~\ref{Fig_individual}.
 Thus, the ultimate precision limit of the global estimation takes the advantages over the local estimation in $\mathrm{SU}(2)$ dynamics. It is important for considered the operators, hence, the approximation is made, see Appendix~\ref{Sec:simultaneous}.  However, when we consider high-dimensional spin and spin-1/2, one can be clear from Fig.~\ref{Fig_individual} that the precision of the parameter estimation is greatly improved for high-dimensional spin  Hamiltonian (spin-7/2 with three-qubit complete graph state). Therefore, we could obtain the precision improvement when the global parameters are estimated with three-qubit complete graph states.

\section{precision-enhancement optimal global estimation}

Now we are in a position to derive the desired bounds through the $\mathrm{SU}(N)$. We start to quantify the entangled degree of the graph state by the average quantum Fisher information. The Hamiltonian operators are the generators of the $\mathrm{SU}(N)$. Then, we show a precision-enhancement scheme with the graph state.

\subsection{Entanglement scaling with QFI}
%======================figure================================================
\begin{figure}[tp]
\includegraphics[width=8.5cm]{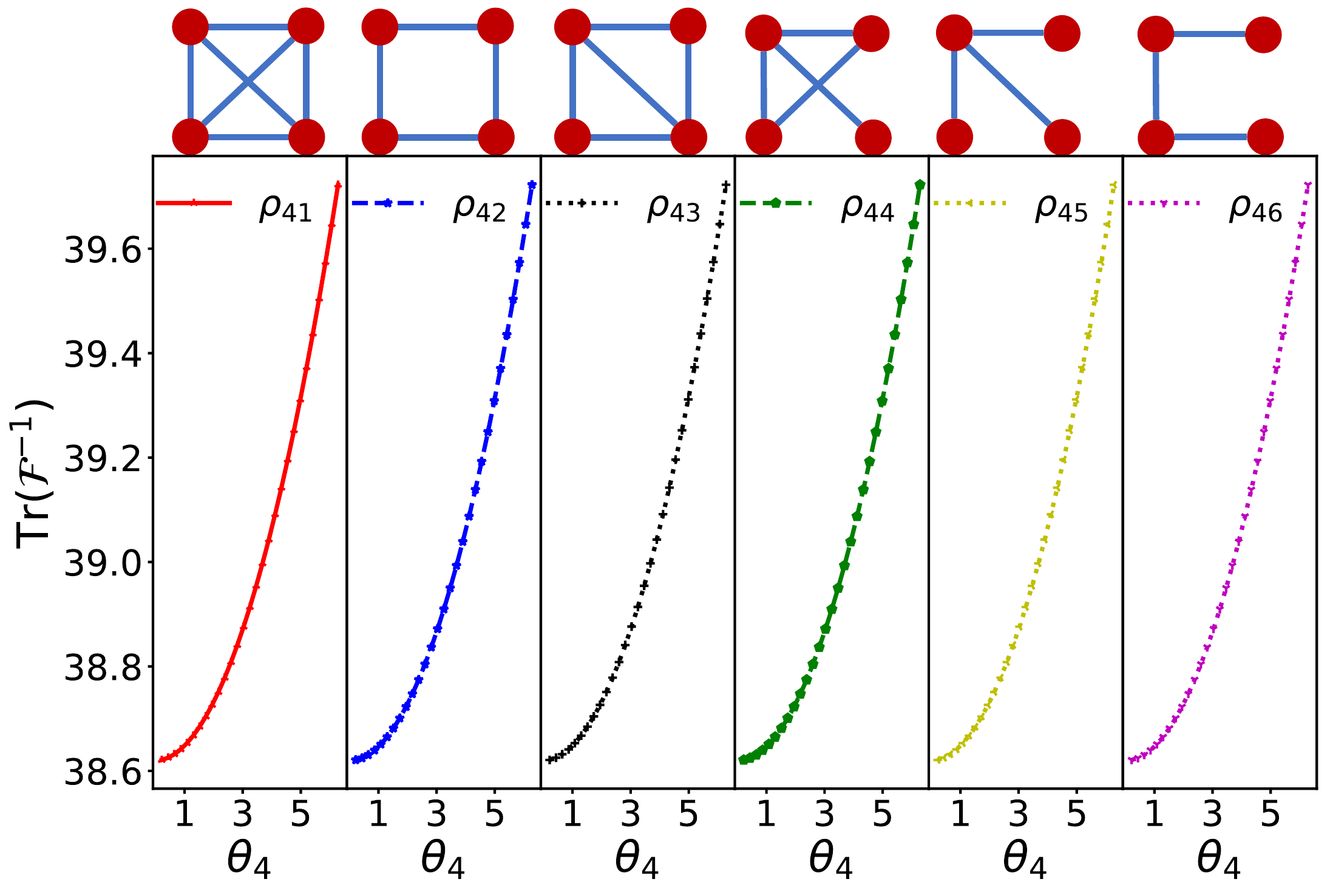} \centering
\caption{(color online). Plot for estimation of the four-qubit graph state with parameters $\theta_4$, and $\theta_1=\theta_2= \theta_3=10^{-2} $. We show the total variance for the six different graph states as shown in inset graph with the different graph structures $\rho_{41}, ~\rho_{42},~\rho_{43},~\rho_{44},~\rho_{45},~\rho_{46} $. For  this  case, the $F_{\text{ave}}$ is the same for the six different graph state with the $\mathrm{SU}(2^n)|_{n=4}$. }\label{Fig_G4_QFI_means}
\end{figure}
%===================================================

Having established the quantum network, we now characterize these entanglement \cite{Srensen2001,Ghne2005}. To quantify two-particle entanglement, we could make rescalingd concurrence $C$ \cite{Wootters1998}.
To quantify multiparticle entanglement, we use the averaged quantum Fisher information (QFI) \cite{Hyllus2012,Toth2012,Lee2014}
\begin{eqnarray}
  F_{\text{ave}}=\frac{4}{3}(\delta^2 J_x+\delta^2 J_y+\delta^2 J_z),
\end{eqnarray}
where $J_{x,y,z}= \frac{1}{2}\sum_{j=1}^n\sigma^{x,y,z}_j$ are the collecter operators on $\mathrm{SU}(2)$ and $\delta^2 J_{x, y, z}=\langle J_{x, y, z}^2 \rangle -\langle J_{x, y, z} \rangle^2 $ are the variance of the operators $J_{x, y, z}$. The graph state has $\langle J_{x, y,x}\rangle =0 $, this reason has been proved in the Appendix~\ref{sec:sigma}. In fact, the average QFI takes the maximum possible value,
$F_{\text{ave}}=(n^2+2n)/3$,
when the quantum states are the complete graph states, i.e. the four-qubit complete graph state is shown in Fig.~\ref{Fig:Est_scheme}(a).
Meaning that the complete graph states are full $n$-particle entangled. In general, the arbitrary graph states for which $\mathrm{Tr}(\rho_0 J_k)=0$ for all $k= x, y, z$ saturated
\begin{equation}
  F_{\text{ave}} \leq  \frac{n^2+2n}{3}.
\end{equation}

For the complete graph states, the dynamics parameterization reason $U_{\theta}=\mathrm{e}^{-\frac{i}{2}\theta \sum_{j}^n \sigma_j^{x,y,z}}$ on the $\mathrm{SU}(2)$, the generators of the $\mathrm{SU}(2)$ is the Pauli $X$, the QFI can be written as $F=n$, and the generators of the $\mathrm{SU}(2)$ is the Pauli $Y (Z)$, the QFI are $F=n^2~(F=n)$. These results of the $\mathrm{SU}(2)$ is proved with multiparameter estimation in the Appendix~\ref{Sec:individual}. For the single parameter, the QFI is the sum of all matrix elements of QFIM.

For the chain graph state, there is the minimum value for the QFI,  the averaged quantum Fisher information (QFI) is
\begin{equation}
  F_{\text{ave}}=n.
\end{equation}
Therefore, there is a minimum degree of entanglement. The ultimate precision of estimation $\delta^2 \bm{\theta}$ has the minimum value with the complete graph state on the $\mathrm{SU}(2)$.

 Now, we can quantify the multiparticle entanglement $\mathrm{SU}(N)$ by using the averaged quantum Fisher information $F_{\text{ave}}$
\begin{align}
  F_{\text{ave}}=\frac{4}{n}(\delta^2 H_1+\delta^2 H_2+ \cdots + \delta^2 H_n),
\end{align}
where $\delta^2 H_j=\langle H_j^2 \rangle -\langle H_j \rangle^2 $ for all $j \in \{1,2,\cdots, n\}$, and the Hermitian operator $H_j$ is the generator of the $\mathrm{SU}(N)$.
We can find
\begin{align}
  F_{\text{ave}}(\mathrm{SU}(N))\geq F_{\text{ave}}(\mathrm{SU}(2^n)), ~~~\forall j, ~k,
\end{align}
where the $N=2^m$ and $m$ is a positive integer and $m \leq n$. When $N^2-1 \leq n$, $N$ takes the minimum value.
For the $n$-qubit graph state, we can find that the averaged quantum Fisher information satisfies
\begin{align}
  F_{\text{ave}}(\rho_{nj})=F_{\text{ave}}(\rho_{nk}), ~~~\forall j, ~k
\end{align}
with $\mathrm{SU}(2^n)$, where $\rho_{nj}$ is the density matrix of the $n$-qubit graph state, i.e., the four-qubit graph state $\rho_{41}, ~\rho_{42},~ \cdots, ~\rho_{46}$ can be shown as  Fig.~\ref{Fig_G4_QFI_means} and the subscript $n$ represents the qubits number of graph state and the $j$ is the classification of graph state, the four-qubit graph state has the classification $j=1,~2,~3,~4,~5,~6$ in Fig.~\ref{Fig_G4_QFI_means}. We can obtain that any $n$-qubit graph state $\rho_0$, the parameters $\theta_k$ to be estimated are the coefficients of a set of generators $H_k$ for $\mathrm{SU}(2^n)$,
the ultimate precision $\delta^2 \bm{\theta}$ depends on the $F_{\text{ave}}$.
when the global Hamiltonian can be written as $H_j=\lambda_j$, $\lambda_j$ is the generators of the $\mathrm{SU}(2^n)$.
\begin{align}
\mathrm{Tr}(\mathcal{F}^{-1})(\rho_{nj})=\mathrm{Tr}(\mathcal{F}^{-1})(\rho_{nk}), ~~~\forall j, ~k.
\end{align}
In the parameters limit $\theta_k \rightarrow 0$, the ultimate precision of the parameter estimation is the same and has the same the entanglement degree with complete global parameterization process in $\mathrm{SU}(2^n)$  for arbitrary $n$-qubit graph states.

\subsection{Precision-Enhancement Optimal Global Estimation}
%======================figure=======================
\begin{figure}[tp]
\includegraphics[width=8.5cm]{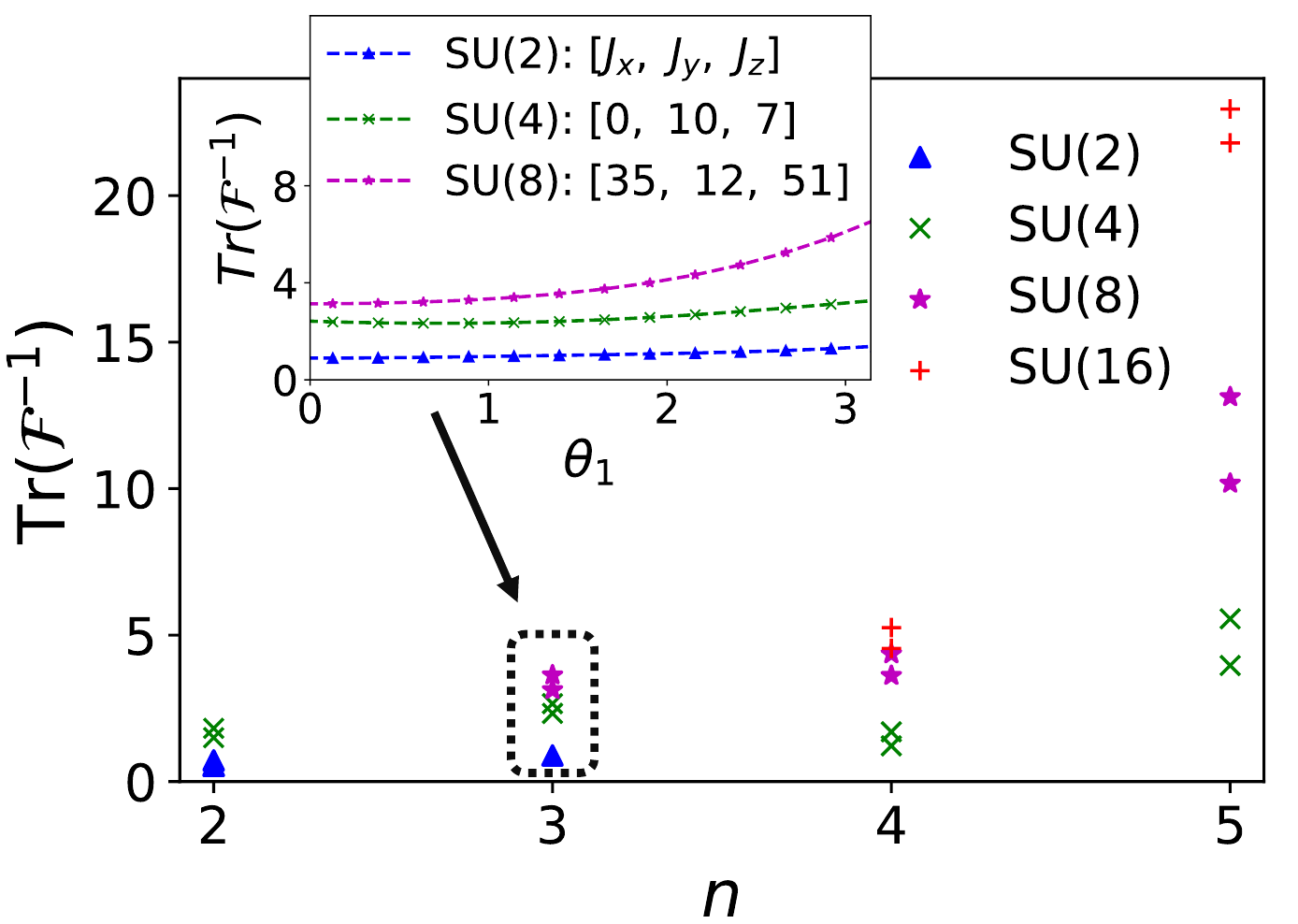} \centering
\caption{(color online). The optimal value of the quantum Cram\'{e}r-Rao bound for complete graph state with the number of the qubits $n=2, ~3, ~4, ~5$. The blue $\triangle$ (green $\times$, purple $\star$,  red+) consider the generators of the $\mathrm{SU}(2)~  (\mathrm{SU}(4), ~\mathrm{SU}(8),~ \mathrm{SU}(16))$. Inset: Total variance for three cases with the three-qubit complete graph state with respect to the parameter value $\theta_1$. The blue dotline represents the generators $J_x, J_y, J_z$ with $\mathrm{SU}(2)$;  the green dotline represents the generators $\lambda_0, \lambda_{10}, \lambda_{7}$ with $\mathrm{SU}(4)$,  the purple dotline represents the generators $\lambda_{35}, \lambda_{12}, \lambda_{51}$ with $\mathrm{SU}(8)$. }\label{Fig_sun_rho}
\end{figure}
%============================================================================

We next show that for the estimation of the general Hamiltonian for the $\mathrm{SU}(N)$. The value of $N$ can have a similar improvement for the precision of parameter estimates of graph states. Furthermore, the advantages, over the best quantum strategies of the estimation improve with the number of $N$.

We now consider any $n$-qubit graph state in which the generators $H_k$ for all $k$ with $\mathrm{SU}(N)$ and the QFIM depends on the parameters $\bm{\theta}$. In the limit $\theta_k \rightarrow 0$,
the quantum Cram\'{e}r-Rao bound satisfies
\begin{align}
  \min (\mathrm{Tr}(\mathcal{F}^{-1})_{\mathrm{SU}(N)}) \leq \min(\mathrm{Tr}(\mathcal{F}^{-1})_{\mathrm{SU}(2^n)}) ,
\end{align}
where the $N=2^m$ and $m$ is a positive integer and $m \leq n$. When $N^2-1 \leq n$, $N$ takes the minimum value.

The analysis shows that the above conclusion is valid. The results are shown in the example shown in Fig.~\ref{Fig_sun_rho}. The ultimate precision can reach the Heisenberg limit for an optimal state, its single-parameter scenario. However, for multiparameter scenarios, that is, to achieve a simultaneous estimation of each parameter, the diagonal elements of the QFIM reach the precision limit of a single parameter. We know that the precision limit is different for different parameterization processes.  In the next section, we show that the graph state is the optimal state for quantum metrology, and show that there is an optimal measurement basis such that classical Fisher information (CFIM) is equal to the QFIM. Thus, the dynamics process will affect the precision of the parameter estimation. From the previous subsection, we can know that the arbitrary $n$-qubit graph state has the same ultimate precision as the graph state in the $\mathrm{SU}(2^n)$ dynamics process. In Fig.~\ref{Fig_sun_rho}, we numerically give the optimal QCRB $\mathrm{Tr}(\mathcal{F}^{-1})$ with the optimal parameters values by the optimal algorithm (Particle swarm optimization, PSO), when the number of particles in the complete graph state is 2, 3, 4, and 5, the corresponding dynamics are $\mathrm{SU}(2)$, $\mathrm{SU}(4)$, $\mathrm{SU}(8)$ and $\mathrm{SU}(16)$, respectively. In the limit $\theta_k  \rightarrow 0$, the inset shows the three-qubit graph state of the QCRB as a function of parameters, corresponding to its dynamics as $\mathrm{SU}(2)$, $\mathrm{SU}(4)$, and $\mathrm{SU}(8)$. We can find the optimal dynamics process as the $\mathrm{SU}(2)$.

\section{Optimal measurement with graph state}
%======================figure================================================
\begin{figure}[tp]
\includegraphics[width=8cm]{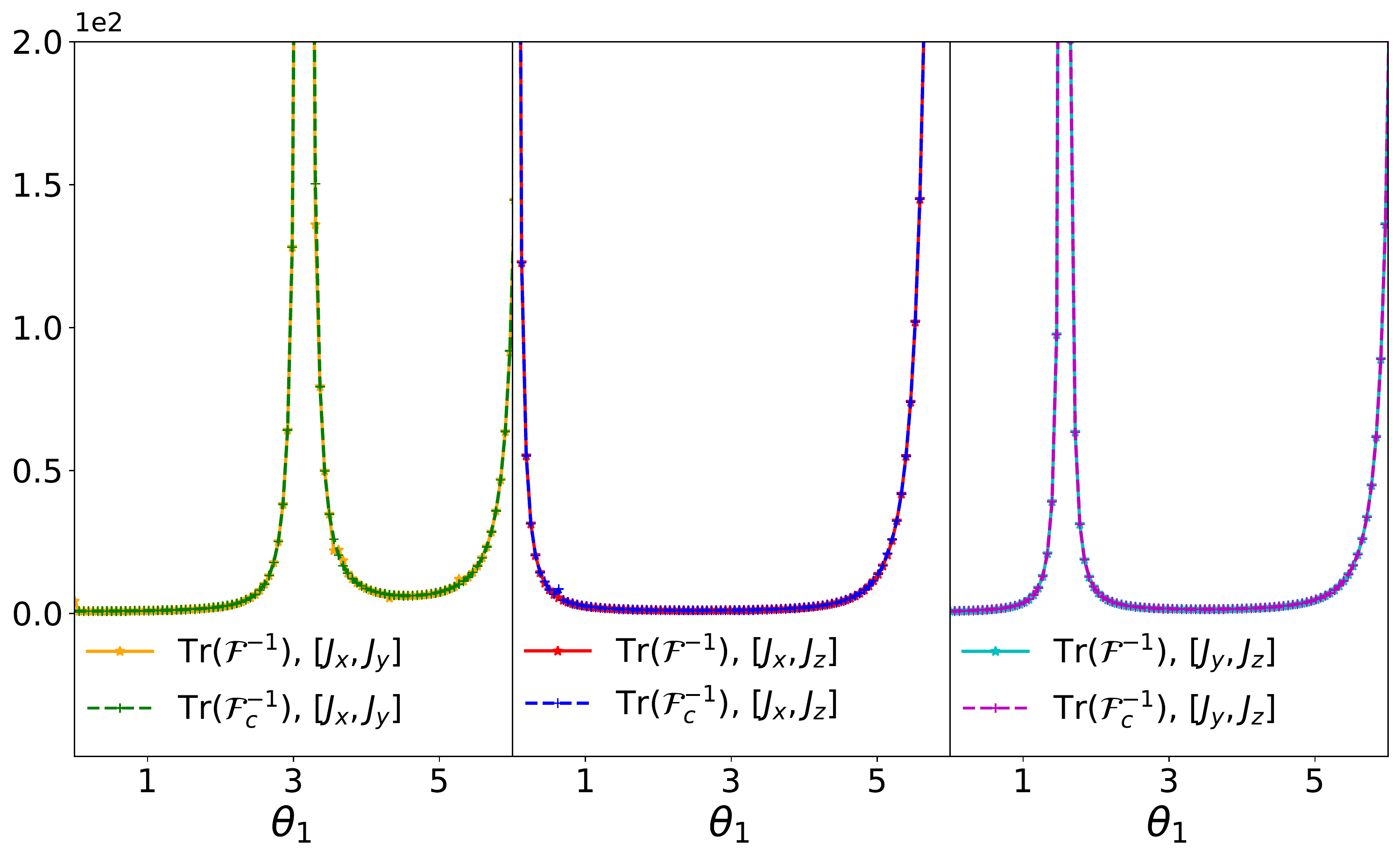}
\centering
\caption{ For the two-qubit graph state, the total variance for the two-parameter estimation on the $\mathrm{SU}(2)$ with the parameter $\theta_1$, and $\theta_2=\theta_1=10^{-3}$. The Hamiltonian operators are the generators of the  $\mathrm{SU}(2)$, $J_x, J_y, J_z$, respectively. As well as the CFIM is obtained with the projection measurement (Bell Measurement). We demonstrate that for the two-qubit graph state, under Bell measurements, the CFIM and QFIM are in perfect anastomotic. Different colored lines represent different Hamiltonians.  The yellow and green lines are the precision limit $\delta^2 \bm{\theta}$ with the Hamiltonian $\bm{H}(\bm{\theta})=\theta_1 J_x + \theta_2 J_y$.
The red and blue lines are the precision limit $\delta^2 \bm{\theta}$ with the Hamiltonian $\bm{H}(\bm{\theta})=\theta_1 J_x + \theta_2 J_z$. The cyan  and purple lines are the precision limit $\delta^2 \bm{\theta}$ with the Hamiltonian $\bm{H}(\bm{\theta})=\theta_1 J_y + \theta_2 J_z$. \label{Fig_opt_measurement}}
\end{figure}
%============================================================================
We have discussed that there is the positive operator-value measurement (POVM) that can attain the multiparameter quantum Cram\'{e}r-Rao bound. For single-parameter estimation, quantum Cram\'er-Rao bound could always be attained with an optimal measurement. However, the quantum Cram\'{e}r-Rao bound for estimation of multiparameter is then not necessarily attainable. When the conditions are satisfied, the optimal measurement always exists and the CFIM saturates the QFIM has long engaged the field of quantum metrology~\cite{Helstrom1976,Holevo1982,Matsumoto2002}. For any unbiased estimators and independent measurement, there are the following inequalities
\begin{align}
\mu \delta^2 \hat{\bm{\theta}} \geq \mathrm{Tr}(\mathcal{F}_c^{-1}) \geq \mathrm{Tr}(\mathcal{F}^{-1}),
\end{align}
where $\mathcal{F}_c$ is the classical Fisher information matrix and the first inequality is the classical and the second inequality the quantum Cram\'{e}r-Rao bound, respectively. $\mu$ represents the number of times that the estimation procedure is repeated.

So a set of optimal measurements can always be found $\mathcal{F}_c=\mathcal{F}$.
For the general unitary parameterization process $U_{\bm{\theta}}$ and arbitrary graph state $\rho_0$,
the sufficient and necessary conditions for the attainability of the quantum Cram\'{e}r-Rao bound are always satisfied \cite{Liu2020}.
\begin{align}
  \mathrm{Tr}(\rho [L_j, L_k])=0, ~\forall j, ~k.
\end{align}
We show that the QFIM is saturated with the Hamiltonian $[H_j, H_k]\neq 0$ for the unitary evolution in Appendix~\ref{sec:Proof_QCRB}. (1) The QFIM is the full rank, (2) the expectation value of the operators $\mathcal{H}$ vanisher for all pairs, i.e., $\mathrm{Im}(\mathrm{Tr}(\mathcal{H}_j \mathcal{H}_k \rho_0))=0$.  The above condition can be equivalent to a sufficient and necessary condition with the operator $\mathcal{H}_j$ and $\mathcal{H}_k$,
\begin{align}
  \mathrm{Tr}(\rho_0 [\mathcal{H}_j, \mathcal{H}_k])=0, ~ \forall j, ~k.
\end{align}
We can consider positive-operator-value measurement sets in which the element is a projection onto the probe state $\rho_{\bm{\theta}}=U_{\bm{\theta}}\rho_0 U^\dagger_{\bm{\theta}}$.
We can find that the attainable set of the measurement and that the classical Fisher information matrix is equal to the QFIM.
Hence, we can find projective measurement orthogonal to the probe state $\rho_{\bm{\theta}}$, which saturates the QFIM. The CFIM $\mathcal{F}_c$ depends on the probe state $\rho_{\bm{\theta}}$.
According to ref. \cite{Pezz2017}, a set of optimal measurements can be found for the arbitrary $n$-qubit graph state.

For the two-qubit graph state, the Bell measurement is the optimal measurement,
\begin{align*}
  |\psi_1\rangle=\frac{|00\rangle+|11\rangle}{
  \sqrt{2}}, ~|\psi_2\rangle=\frac{|00\rangle-|11\rangle}{
  \sqrt{2}}, \\
  |\phi_1\rangle=\frac{|01\rangle+|10\rangle}{
  \sqrt{2}}, ~|\phi_2\rangle=\frac{|01\rangle-|10\rangle}{
  \sqrt{2}}.
\end{align*}
In the parameters limit $\theta_j \rightarrow 0$, the Hamiltonian operator is $\bm{H}(\bm{\theta})=\theta_1 J_x +\theta_2 J_z$ which is the generators of the $\mathrm{SU}(2)$. Based on  the  probability  distribution  it  is  straightforward  to  calculate  the CFIM as
\begin{equation}
\mathcal{F}_c=\left(\begin{array}{cc}
4\langle J_x^2\rangle & 0  \\
0 & 4\langle J_z^2\rangle
\end{array}\right).
\end{equation}
In Fig.~\ref{Fig_opt_measurement},  the CFIM is the same as the QFIM with the generators of the $\mathrm{SU}(2)$. The quantum Cram\'{e}r-Rao bound is thus asymptotically saturable in this case. We found that the graph state is the optimal quantum state in quantum metrology and there is a set of the optimal measurement which could saturate the quantum Cram\'{e}r-Rao bound for an arbitrary generator of the phase encoding on the dynamics $\mathrm{SU}(N)$.

\section{Conclusion}
The graph state is the essential quantum resource of quantum information processing. Shettell et al. proposed that the graph state is an important quantum resource for quantum precision measurement \cite{Shettell2020}. By constructing the graph state cluster, the precision limit of its single parameter estimation could attain the Heisenberg limit. Furthermore, we investigate multiparameter estimation on graph state for quantum metrology. We developed a method associated with the dynamics process of the $\mathrm{SU}(N)$, which is an encoding phase to investigate the quantum multiparameter estimation. The Hamiltonian operators are the generators of the $\mathrm{SU}(N)$.

We discuss the two scenarios of local estimation and global estimation, respectively. When the generators of $\mathrm{SU}(N)$ are commutative with each other $[H_j, H_k]=0$, the QFIM at this time is parameter independent, and when the Hamiltonian operator is not composed of the same generator tensor of the $\mathrm{SU}(N)$, the QFIM is invertible. However, when the Hermitian operators between generators are not commutative $[H_j, H_k] \neq 0$, the QFIM at this time is parameter-dependent. Under the parameter limit $\theta_j \rightarrow 0$, we give the expression of its QFIM and explore whether the precision of the global parameter estimation is higher than local parameter estimation when the unitary parameterization process is the $\mathrm{SU}(2)$ dynamics.
In addition, for the $n$-qubit graph state, we give a precision enhancement scheme. In $\mathrm{SU}(N)$ dynamics, when the value of $N$ is smaller, the precision is higher, and the maximum value of $N$ is equal to $2^n$. When $N=2^n$, the $n$-qubit graph states have the same entanglement degree with the different graph structure, that is, the averaged quantum Fisher information $F_{\text{ave}}$ have the same values. At this time, the precision limit of the $n$-parameter estimation is exactly the same for different graph states.
Finally, we show that the graph state is the optimal quantum state in quantum metrology, and we prove that for any unitary parameter process, a set of optimal measurement basis can be found, making the precision limit could attain the quantum Cram\'{e}r-Rao bound.

This shows that quantum graph states are good resources for the robust quantum metrology.  From this perspective, the graph state, as the quantum maximally entangled pure state is the natural choice for integrating quantum sensing into the future quantum network, which is of great significance to the study of quantum computing and quantum information processing.

\begin{acknowledgments}
  The authors would like to thank Prof.\,Jing Liu for helpful discussion.
\end{acknowledgments}

%\bibliography{graphmetro_manucsriptNotes}

\begin{thebibliography}{1}

\bibitem{West2001}
D. B. West, \emph{Introduction to Graph Theory},
(Prentice Hall, Upper Saddle River, NJ, 2001).

\bibitem{Diestel2000}
R. Diestel, \emph{Graph Theory}
(Prentice Hall, Upper Saddle River, NJ, 2001).

\bibitem{Raussendorf2003}
R. Raussendorf, D. E. Browne, and H. J. Briegel,
Measurement-based quantum computation on cluster states,
\href{https://doi.org/10.1103/PhysRevA.68.022312}
{Phys. Rev. A \textbf{68}, 022312 (2003).}

\bibitem{Aschauer2003}
W. Dür, H. Aschauer, and H. J. Briegel, Multiparticle Entanglement Purification for Graph States,
 \href{https://doi.org/10.1103/PhysRevLett.91.107903}
{Phys. Rev. Lett. \textbf{91}, 107903(2003).}

  \bibitem{Grassl2002}
  M. Grassl, A. Klappenecker, and M. Rötteler, Graphs, Quadratic Forms, and Quantum Codes, in Proceedings of  the  2002  IEEE  International  Symposium  on  Information Theory,
  \href{https://doi.org/10.1109/ISIT.2002.1023317}
  { Lausanne, Switzerland, 2002, p. 45.}

  \bibitem{Briegel2001}
  H.-J. Briegel and  R. Raussendorf,
  Persistent Entanglement in Arrays of Interacting Particles,
  \href{https://doi.org/10.1103/PhysRevLett.86.910}
  {Phys. Rev. Lett. \textbf{86}, 910 (2001).}

  \bibitem{Hein2004}
  M. Hein, J. Eisert and H.-J. Briegel
  Persistent Entanglement in Arrays of Interacting Particles,
  \href{https://doi.org/10.1103/PhysRevLett.86.910}
  {Phys. Rev. A \textbf{69}, 062311 (2004).}

  \bibitem{Aschauer2005}
   H. Aschauer, W. Dür, and H.-J. Briegel,
  Multiparticle entanglement purification for two-colorable graph states,
  \href{https://doi.org/10.1103/PhysRevA.71.012319}
  {Phys. Rev. A \textbf{71}, 012319 (2005).}

  \bibitem{NEST2004}
   M. Van den Nest, J. Dehaene and B. De Moor,
  Graphical description of the action of local Clifford transformations on graph states,
  \href{https://doi.org/10.1103/PhysRevA.69.022316}
  {Phys. Rev. A \textbf{69}, 022316 (2004).}


  \bibitem{Gottesman1997}
  D. Gottesman,
  Stabilizer Codes and Quantum Error Correction, PhD thesis, CalTech,Pasadena (1997).

  \bibitem{Schlingemann2002}
  D. Schlingemann and R. F. Werner,
  Quantum error-correcting codes associated with graphs,
  \href{https://doi.org/10.1103/PhysRevA.65.012308}
  {Phys. Rev. A \textbf{65}, 012308 (2001)}

  \bibitem{Raussendorf2001}
  R.  Raussendorf  and  H. -J.  Briegel,
  A One-Way Quantum Computer,
\href{https://doi.org/10.1103/PhysRevLett.86.5188}
  {Phys. Rev. Lett. \textbf{86}, 5188 (2001)}

  \bibitem{Shettell2020}
  N. Shettell and D. Markham, Graph States as a Resource for Quantum Metrology,
\href{https://journals.aps.org/prl/abstract/10.1103/PhysRevLett.124.110502}
  {Phys. Rev. Lett. \textbf{124}, 110502 (2020)}

  \bibitem{Liu2020}
  J. Liu, H. Yuan, X.-M. Lu, and X. Wang,
  Quantum Fisher information matrix and multiparameter estimation,
  \href{https://iopscience.iop.org/article/10.1088/1751-8121/ab5d4d}
  {J. Phys. A: Math. Theor. \textbf{53}, 023001 (2020).}

  \bibitem{Szczykulska2016}
  M. Szczykulska, T. Baumgratz and A. Datta,
  Multi-parameter quantum metrology,
  \href{https://doi.org/10.1080/23746149.2016.1230476}
  {Adv. Phys. X \textbf{1}, 621 (2016).}

  \bibitem{Liu2022}
  M. Zhang, H-M. Yu, H. Yuan, X. G. Wang, R. Demkowicz-Dobrza\ifmmode \acute{n}\else \'{n}\fi{}ski, and J. Liu,
  QuanEstimation: An open-source toolkit for quantum parameter estimation,
  \href{https://doi/10.1103/PhysRevResearch.4.043057}
  {Phys. Rev. Research \textbf{4}, 043057 (2022).}

\bibitem{Albare2020}
F. Albarelli, M. Barbieri, M. G. Genonie, and I. Giananifc, A perspective on multiparameter quantum metrology: From theoretical tools to applications in quantum imaging,
\href{https://doi.org/10.1016/j.physleta.2020.126311}
{Phys. Lett. A \textbf{384}, 126311 (2020).}

\bibitem{Apellaniz2016}
G. T\'oth and I. Apellaniz, Quantum metrology from aquantum information science perspective,
\href{https://doi.org/10.1088/1751-8113/47/42/424006}
{J. Phys. A: Math. Theor. \textbf{47}, 424006  (2014).}

\bibitem{Giovannetti2006}
V. Giovannetti, S. Lloyd, and L. Maccone, Quantum Metrology,
\href{https://doi.org/10.1103/PhysRevLett.96.010401}
{Phys. Rev. Lett. \textbf{96}, 010401 (2006).}

\bibitem{Giovannetti2004}
V. Giovannetti, S. Lloyd, and L. Maccone, Quantum-enhanced measurements: beating the standard quantumlimit,
\href{https://doi.org/10.1126/science.1104149}
{Science \textbf{306},1330 (2004).}

  \bibitem{Giovannetti2011}
  V. Giovannetti, S. Lloyd, and L. Maccone, Advances in quantum metrology,
  \href{https://doi.org/10.1038/nphoton.2011.35}
  {Nat. Photonics \textbf{5},222 (2011).}

\bibitem{Demkowicz2015}
R. Demkowicz-Dobrzanski, M. Jarzyna, and J. Kolodynski, Chapter Four-Quantum Limits in Optical Interferometry
\href{https://doi.org/10.1016/bs.po.2015.02.003}
  {Prog. Opt. \textbf{60}, 345 (2015).}

\bibitem{Demkowicz2012}
R. Demkowicz-Dobrza\'nski, J. Kolody\'nski, and M. Guţ\"a, The elusive Heisenberg limit in quantum-enhanced metrology,
  \href{https://doi.org/10.1038/ncomms2067}
  {Nat. Commun.\textbf{3}, 1063 (2012).}

\bibitem{Baumgratz2016}
  T. Baumgratz and A. Datta, Quantum Enhanced Estimation of a Multidimensional Field,
  \href{https://doi.org/10.1103/PhysRevLett.116.030801}{Phys. Rev. Lett.\textbf{116}, 030801 (2016)}

\bibitem{Hall2012}
L. T. Hall, G. C. G. Beart, E. A. Thomas, D. A. Simpson, L. P. McGuinness, J. H. Cole, J. H. Manton, R. E. Scholten, F. Jelezko, J. Wrachtrupet al., High spatial and temporal resolution wide-field imaging of neuron activity using quantum NV-diamond,
\href{https://doi.org/10.1038/srep00401}
{ Sci. Rep.\textbf{2}, 401 (2012).}

\bibitem{Seo2007}
M. A. Seo, A. J. L. Adam, J. H. Kang, J. W. Lee, S. C. Jeoung, Q. H. Park, P. C. M. Planken, and D. S. Kim, Fourier-transform terahertz near-field imaging of one-dimensional slitarrays: Mapping of electric-field-, magnetic-field-, and Poynting vectors,
\href{https://doi.org/10.1364/oe.15.011781}
{Opt. Express \textbf{15}, 11781 (2007).}

\bibitem{Steinert2010}
S. Steinert, F. Dolde, P. Neumann, A. Aird, B. Naydenov, G.Balasubramanian, F. Jelezko, and J. Wrachtrup, Highsensitivity magnetic imaging using an array of spins indiamond,
\href{https://doi.org/10.1063/1.3385689}
{Rev. Sci. Instrum.\textbf{81}, 043705 (2010).}

\bibitem{Pham2011}
L. M. Pham, D. Le Sage, P. L. Stanwix, T. K. Yeung, D.Glenn, A. Trifonov, P. Cappellaro, P. R. Hemmer, M. D. Lukin, H. Parket et al., Magnetic field imaging with nitrogen-vacancy ensembles,
\href{https://doi.org/10.1088/1367-2630/13/4/045021}
{New J. Phys. \textbf{13},  045021 (2011).}

\bibitem{Liu2016}
J. Liu, X.-M. Lu, Z. Sun, and X. Wang, Quantum multi-parameter metrology with generalized entangled coherent state,
\href{https://doi.org/10.1088/1751-8113/49/11/115302}
{ J. Phys. A: Math. Theor. \textbf{49}, 115302 (2016).}

\bibitem{Humphreys2013}
P. C. Humphreys, M. Barbieri, A. Datta, and I. A. Walmsley, Quantum Enhanced Multiple Phase Estimation,
\href{https://doi.org/10.1103/PhysRevLett.111.070403}
{Phys. Rev. Lett.  \textbf{111}, 070403 (2013).}

\bibitem{Ciampini2016}
M. A. Ciampini, N. Spagnolo, C. Vitelli, L. Pezz\'e, A. Smerzi, and F. Sciarrino, Quantum-enhanced multiparameter estimation in multiarm interferometers,
\href{https://doi.org/10.1038/srep28881}
{Sci.Rep. \textbf{6}, 28881 (2016).}

\bibitem{Yue2014}
J.-D. Yuecoherent state, and H. Fan, Quantum-enhanced metrology for multiple phase estimation with noise,
\href{https://doi.org/10.1038/srep05933}
{Sci. Rep. \textbf{4}, 5933 (2014).}

\bibitem{Knott2016}
P. A. Knott, T. J. Proctor, A. J. Hayes, J. F. Ralph, P. Kok, andJ. A. Dunningham, Local versus global strategies in multi-parameter estimation,
\href{https://doi.org/10.1103/PhysRevA.94.062312}
{Phys. Rev. A \textbf{94}, 062312 (2016).}

\bibitem{Gagatsos2016}
C. N. Gagatsos, D. Branford, and A. Datta, Gaussian systems for quantum-enhanced multiple phase estimation,
\href{https://doi.org/10.1103/PhysRevA.94.042342}
{Phys. Rev. A \textbf{94}, 042342 (2016).}

\bibitem{Zhang2017}
L. Zhang and K. W. C. Chan, Quantum multiparameter estimation with generalized balanced multimode noon-likestates,
\href{https://doi.org/10.1103/PhysRevA.95.032321}
{Phys. Rev. A \textbf{95}, 032321 (2017).}

\bibitem{Komar2014}
P. Komar, E. M. Kessler, M. Bishof, L. Jiang, A. S. Sørensen, J. Ye, and M. D. Lukin, A  quantum network of clocks,
\href{https://doi.org/10.1038/nphys3000}
{Nat. Phys. \textbf{10}, 582 (2014).}

\bibitem{Proctor2018}
T. J. Proctor, P. A. Knott, and J. A. Dunningham,
Multiparameter Estimation in Networked Quantum Sensors,
\href{https://journals.aps.org/prl/abstract/10.1103/PhysRevLett.120.080501}
{Phys. Rev. Lett. \textbf{120}, 080501 (2018).}

\bibitem{Gessner2018}
M. Gessner, L. Pezze, and A. Smerzi, Sensitivity Bounds for Multiparameter Quantum Metrology,
\href{https://doi.org/10.1103/PhysRevLett.121.130503}{Phys. Rev. Lett.\textbf{121}, 130503 (2018)}

\bibitem{Yuan2016}
H. D. Yuan,
Sequential Feedback Scheme Outperforms the Parallel Scheme for Hamiltonian Parameter Estimation,
\href{https://doi.org/10.1103/PhysRevLett.117.160801}{Phys. Rev. Lett.\textbf{117}, 160801 (2016)}

\bibitem{Imai2007}
H. Imai and A. Fujiwara, Geometry of optimal estimation scheme for SU(D) channels,
\href{http://doi.org/10.1088/1751-8113/40/16/009}
{J. Phys. A \textbf{40}, 4391 (2007)}

\bibitem{Goldberg2021}
A.  Z.  Goldberg,  L.  L.  Sánchez-Soto,  and  H.  Ferretti, Intrinsic Sensitivity Limits for Multiparameter Quantum Metrology,
\href{https://doi.org/10.1103/PhysRevLett.127.110501}
{Phys.Rev.Lett.  \textbf{127}, 110501 (2021)}

\bibitem{Pezz2017}
L. Pezz\'{e}, M. A. Ciampini, N. Spagnolo,P. C. Humphreys, A. Datta, I. A. Walmsley,M. Barbieri, F. Sciarrino,  and A. Smerzi, Optimal Measurements for Simultaneous Quantum Estimation of Multiple Phases,
\href{https://doi.org/10.1103/PhysRevLett.119.130504}
{Phys. Rev. Lett.  \textbf{119}, 130504 (2017).}

\bibitem{Chen2022}
H. Z. Chen, Y, Chen. and H. D. Yuan,, Information Geometry under Hierarchical Quantum Measurement,
\href{https://doi.org/10.1103/PhysRevLett.128.250502}
{Phys. Rev. Lett. \textbf{128}, 250502 (2022).}

\bibitem{Albarelli2020}
F.  Albarelli,  M.  Barbieri,  M.  G.  Genoni,  and  I.  Gianani,  A perspective on multiparameter quantum metrology: From theoretical tools to applications in quantum imaging,
\href{https://doi.org/10.1016/j.physleta.2020.126311}
{Phys. Lett. A \textbf{384}, 126311 (2020).}

\bibitem{Rubio2020}
J.  Rubio,  P.  A.  Knott,  T.  J.  Proctor,  and  J.  A.  Dunningham, Quantum  sensing  networks  for  the  estimation  of  linear  functions,
\href{https://doi.org/0.1088/1751-8121/ab9d46}
{ J. Phys. A: Math. Theor. \textbf{53}, 344001 (2020).}

\bibitem{Gross2021}
J.  A.  Gross  and  C.  M.  Caves,  One  from  many:  Estimating  afunction  of  many  parameters,
\href{https://doi.org/10.1088/1751-8121/abb9ed}
{ J. Phys. A: Math. Theor. \textbf{54}, 014001 (2021).}

\bibitem{Triggiani2019}
D.  Triggiani,  P.  Facchi,  and  V.  Tamma,  Heisenberg scaling precision in the estimation of functions of parameters in linear optical networks,
\href{https://doi.org/10.1103/PhysRevA.104.062603}
{Phys.Rev. A  \textbf{104}, 062603 (2021).}

\bibitem{Helstrom1976}
C. W. Helstrom, \emph{Quantum Detection and Estimation Theory} (New York: Academic, 1976).

\bibitem{Holevo1982}
A. S. Holevo, \emph{Probabilistic and Statistical Aspects of Quantum Theory}
(Amsterdam: North-Holland, 1982).

\bibitem{Matsumoto2002}
 K. Matsumoto, A new approach to the Cram\'{e}r-Rao-type bound of the pure-state model,
\href{https://doi.org/10.1088/0305-4470/35/13/307}
{J. Phys. A: Math. Gen.  \textbf{35}, 3111 (2002).}

\bibitem{Zhu2015}
H. Zhu, Information complementarity: A new paradigm fordecoding quantum incompatibility,
\href{https://doi.org/ 10.1038/srep14317}
{Sci. Rep.\textbf{5}, 14317 (2015).}

\bibitem{Heinosaari2016}
T. Heinosaari, T. Miyadera, and M. Ziman, An invitation to quantum incompatibility,
\href{https://doi.org/10.1088/1751-8113/49/12/123001}
{J. Phys. A: Math. Theor. \textbf{49}, 123001 (2016).}

\bibitem{Ragy2016}
S. Ragy, M. Jarzyna, and R. Demkowicz-Dobrzański, Compatibility in multiparameter quantum metrology,
\href{https://doi.org/10.1103/PhysRevA.94.052108}
{Phys.Rev. A \textbf{94}, 052108 (2016).}


\bibitem{Omolo2018}
 J. A. Omolo, Determining SU(N) symmetry group generators,
\href{https://doi.org/10.13140/RG.2.2.17815.21920}
{ ResearchGate-Preprint, (2018).}


\bibitem{Haber2021}
H. Haber, Useful relations among the generators in the defining and adjoint  representations of SU(N),
\href{https://doi.org/10.21468/SciPostPhysLectNotes.21}
{ SciPost Phys. Lect. Notes, \textbf{21},  (2021).}


\bibitem{Srensen2001}
A. S. Sørensen and K. Mølmer, Entanglement and Extreme Spin Squeezing,
\href{https://doi.org/10.1103/PhysRevLett.86.4431}
{Phys. Rev. Lett. \textbf{86}, 4431 (2001)}

\bibitem{Ghne2005}
O. Gühne, G. Tóth, and H. J. Briegel, Multipartite entanglement in spin chains,
\href{https://doi.org/10.1103/PhysRevLett.86.4431}
{New J. Phys. \textbf{7}, 229 (2005)}

\bibitem{Wootters1998}
W. K. Wootters, Entanglement of Formation of an Arbitrary State of Two Qubits,
\href{https://doi.org/10.1103/PhysRevLett.80.2245}
{Phys. Rev. Lett. \textbf{80}, 2245 (1998)}

\bibitem{Hyllus2012}
P. Hyllus, W. Laskowski, R. Krischek, C. Schwemmer,W. Wieczorek, H. Weinfurter, L. Pezzé, and A. Smerzi, Fisher information and multiparticle entanglement,
\href{https://doi.org/10.1103/PhysRevA.85.022321}
{Phys. Rev. A \textbf{85}, 022321 (2012)}

\bibitem{Toth2012}
G. T\'oth, Multipartite entanglement and high-precision metrology,
\href{https://doi.org/10.1103/PhysRevA.85.022322}
{Phys. Rev. A \textbf{85}, 022322 (2012)}

\bibitem{Lee2014}
 T. E. Lee, F. Reiter, and N. Moiseyev, Entanglement and Spin Squeezing in Non-Hermitian Phase Transitions,
\href{https://doi.org/10.1103/PhysRevLett.113.250401}
{Phys. Rev. Lett. \textbf{113}, 250401 (2014)}

\end{thebibliography}

%\clearpage{}
%\widetext \setcounter{figure}{0}\global\long\def\thefigure{S\arabic{figure}}
%\setcounter{equation}{0} \global\long\def\theequation{S\arabic{equation}}

%\begin{center}
%\textbf{\Large{}{}Supplemental Material }{\Large{} }
%\par\end{center}{\Large \par}

\appendix

\section{Calculation of the QFIM with individual estimation} \label{Sec:individual}

\begin{figure}[tp]
\centering\includegraphics[width=8.5cm]{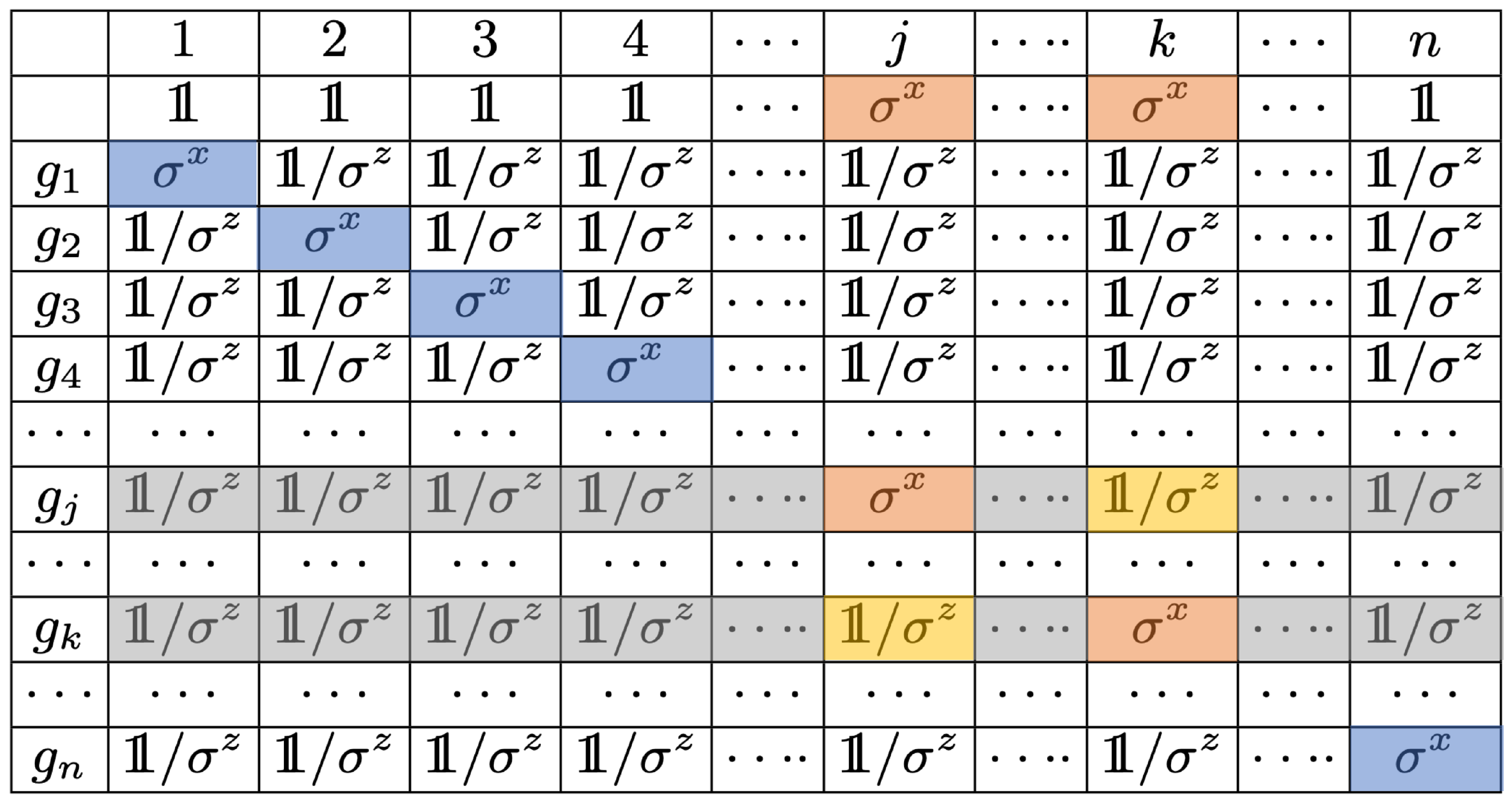}
\caption{The structure of all the stabilizers. The $i$th position
of $g_i$ is $\sigma^x$ (blue box) and other positions are $\openone$ or $\sigma^z$.
\label{fig:apx_gi} }
\end{figure}

The density matrix of the graph state we consider is
\begin{equation} \label{eq:gra_sta}
\rho_{0}= \prod_{i=1}^{n}\frac{1}{2^n}(g_{i}+\openone^{\otimes n}),
\end{equation}
where $\openone$ is a 2-dimensional identity matrix and $g_i$ is the stabilizer
of $i$th qubit, which is of the form
\begin{equation}
g_{i}=\sigma^x_{i} \bigotimes_{j \in \mathcal{N}(i)} \sigma^z_{j},
\end{equation}
with $\mathcal{N}(i)$ the neighborhood of the $i$th qubit and $\sigma^{x(z)}_i$ is the Pauli $X
(Z)$ matrix for the $i$th qubit. The structure of $g_i$ is illustrated in Fig.~\ref{fig:apx_gi}.
The $i$th position in the tensor for $g_i$ is $\sigma^x$, and $\openone$ or $\sigma^z$ for
other positions.

Firstly, the parameterization process is performed for $\mathrm{SU}(2)$ via the operator
\begin{equation}
U_{\bm{\theta}}= \exp \left(- \frac{i}{2} \sum_{j=1}^{n} \theta_{j}\sigma^k_{j}\right)
=\prod_{j=1}^{n}\exp\left(-\frac{i}{2}\theta_{j}\sigma^k_{j}\right),
\end{equation}
where $\bm{\theta}=(\theta_1, \theta_2, \cdots,\theta_n)$ is a vector of unknown parameters, and $k= x, y, z$, $\sigma^k$ is the Pauli $X (Y, Z)$ operator. Next, take Pauli $X$ as an example to find the QFIM of the multiparameter case.
For such a process, the entry of QFIM $\mathcal{F}_{jk}$ (short for $\mathcal{F}_{\theta_{j},\theta_{k}}$)
can be expressed by
\begin{equation}
\mathcal{F}_{jk}=\mathrm{Tr}\left(\sigma^x_{j}\sigma^x_{k}\rho_0\right)
-\mathrm{Tr}\left(\sigma^x_{j}\rho_0\right)\mathrm{Tr}\left(\sigma^x_{k}\rho_0\right).
\end{equation}

%============================= Figure ==========================================
%===============================================================================

One should notice that $\rho_0$ can be rewritten into
\begin{eqnarray}\label{eq:gra_stabi}
2^n \rho_0 &=&\openone+\sum_{i}g_i+\sum_{i<j}g_i g_j+\cdots \nonumber \\
& & +\!\!\!\!\sum_{i<j<k<\cdots <l}\!\!\!\!g_i g_j g_k\cdots g_l \!+\cdots+g_1 g_2 \cdots g_n.
\end{eqnarray}
For the term $\sigma^x_j \rho_0$, the $j$th position of $\sigma^x_j g_j$ is $\openone$.
However, this position of any other $g_i$ is $\openone$ or $\sigma^z$, which means
this position of $\sum_{i<j<k<\cdots <l}g_i g_j g_k\cdots g_l$ is $\sigma^x$ or $\sigma_y$,
a traceless matrix. Furthermore, since the trace of a tensor is zero if any position in
this tensor is traceless, one can find that $\mathrm{Tr}(\sigma^x_j \rho_0)=0$ for
any $j$.

With respect to the term $\mathrm{Tr}(\sigma^x_j \sigma^x_k\rho_0)$, the similar
analysis tells us that the only possible non-zero term is $\mathrm{Tr}(\sigma^x_j\sigma^x_k g_j g_k)$,
namely,
\begin{equation}
\mathrm{Tr}(\sigma^x_j \sigma^x_k\rho_0)=\mathrm{Tr}(\sigma^x_j\sigma^x_k g_j g_k).
\end{equation}
In the case that $j=k$, the equation above reduces to $\mathrm{Tr}(g_j^2)$, which
equals to 1 since $g_i^2=\openone$ (see Fig.~\ref{fig:apx_gi}). For $j\neq k$,
the only case that can make above equation non-zero is the multiplication of
$\sigma^x_j g_j$ and $\sigma^x_k g_k$ is $\openone^{\otimes n}$. However, the
$j$th, $k$th positions of $\sigma^x_j g_j$ and $\sigma^x_k g_k$ are $\openone$
(orange boxes in Fig.~\ref{fig:apx_gi}), respectively. To make it happen, the $k$th
position of $\sigma^x_j g_j$ and $j$th position of $\sigma^x_k g_k$ have to be $\openone$
too (yellow boxes in Fig.~\ref{fig:apx_gi}), which means $j$ and $k$ cannot be neighbors.
Next, all the other positions (apart from $j$th, $k$th) of $\sigma^x_j g_j$ and
$\sigma^x_k g_k$ have to be the same, indicating that $j$th and $k$th qubit must
have the same neighborhood, i.e., $\mathcal{N}(j)=\mathcal{N}(k)$. In the mean
time, the condition $\mathcal{N}(j)=\mathcal{N}(k)$ also includes the requirement
that $j$, $k$ are not neighbors as the neighbor qubits cannot share the same
neighborhood as they have each other in the neighborhood. Hence, the only case
that make $\mathrm{Tr}(\sigma^x_j\sigma^x_k g_j g_k)$ ($j\neq k$) non-zero is
that $\mathcal{N}(j)=\mathcal{N}(k)$. The off-diagonal entries of QFIM can then
be expressed by
\begin{equation}
\mathcal{F}_{jk}=\begin{cases}
1,~\mathrm{for}~\mathcal{N}(j)=\mathcal{N}(k),\\
0,~\mathrm{for}~\mathcal{N}(j)\neq\mathcal{N}(k).
\end{cases}
\label{eq:Sup_QFIM_n}
\end{equation}

In the case that the parameter number is less than $n$, the parameterization
the operator can be written as
\begin{equation}
U_{\bm{\theta}}=\exp\left(-\frac{i}{2}\sum_{j=1}^{d}\theta_{j}\sum_{k\in S_{j}}\sigma^x_{k}\right),
\end{equation}
where $S_j$ is the set of vertices that are used to encode $\theta_j$. The QFIM can be written as
\begin{equation}
\mathcal{J}_{jm}=\sum_{k_{j}=S_{j-1}+1}^{S_{j}}\sum_{l_{m}=S_{m-1}+1}^{S_{m}} \mathcal{F}_{k_j, l_m}, ~~~ j\neq m,
\end{equation}
where
\begin{align}
\mathcal{F}_{k_j, l_m}=\mathrm{Tr}\left(\sigma^x_{k_j}\sigma^x_{l_m}\rho_0\right)
-\mathrm{Tr}\left(\sigma^x_{k_j}\rho_0\right)\mathrm{Tr}\left(\sigma^x_{l_m}\rho_0\right).
\end{align}
Therefore, the diagonal element of the QFIM satisfies the value of the QFI in the form of a subgraph $S_{j}$. For $\mathrm{SU}(2)$, we can find that the QFIM is not necessarily invertible.
For Pauli Y case, the off-diagonal entries of QFIM can then be expressed by
\begin{equation}
\mathcal{F}_{jk}=\begin{cases}
1,~\mathrm{for}~\mathcal{N}(j)+k=\mathcal{N}(k)+j,\\
0,~\mathrm{for~others}.
\end{cases}
\label{eq:Sup_QFIM_y}
\end{equation}
And the diagonal entries of QFIM can then be expressed by $\mathcal{F}_{jj}=1$ for all $j$.
For Pauli-$Z$ case, the QFIM can be obtained
\begin{equation}
  \mathcal{F}= \openone_n,
\end{equation}
where $\openone_n$ is the $n$-dimension identity matrix.

Secondly, the parameterization process for $\mathrm{SU}(N)$ is performed via the operator
\begin{equation}
U_{\bm{\theta}}=\exp\left(-i\sum_{j=1}^n \theta_j H_j\right),
\end{equation}
 where $H_j=\frac{1}{2}\sum_j^s\lambda_{j}$, and the Hermitian operators $\lambda_i$ are the generator of the $\mathrm{SU}(N), N\geq 4$, and $[H_j, H_k]=0$ for all $j,k$ with the parameters $\theta_j, \theta_k$. The element of QFIM can be expressed by
\begin{align}
\mathcal{F}_{j, k }=4[\mathrm{Tr}\left(H_j H_k\rho_0\right)-\mathrm{Tr}\left(H_j\rho_0\right)\mathrm{Tr}\left(H_k\rho_0\right)].
\end{align}
We can find that the QFIM is parameter independent, the quantum Cram\'{e}r-Rao bound satisfies
\begin{align}
  \mathrm{Tr}(\mathcal{F}^{-1}) =c,
\end{align}
where $c$ is the constant number.

\section{Proof the Pauli operators on SU(2)}\label{sec:sigma}
%============================ Figure ===========================
\begin{figure}[tp]
\centering\includegraphics[width=8.5cm]{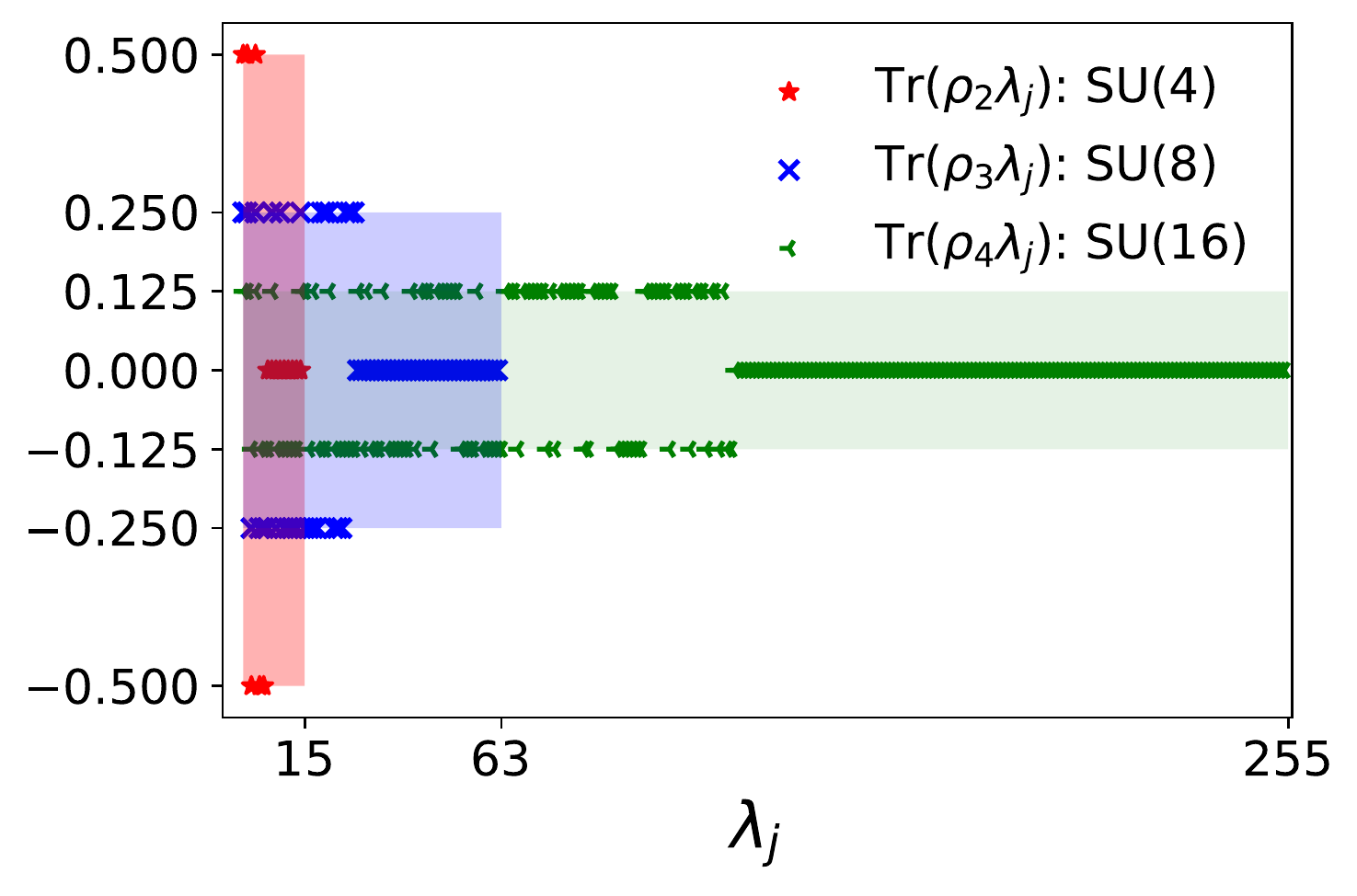}
\caption{Example to demonstrate the average value $\mathrm{Tr}(\rho_0 \lambda_j)$ that when the generator $\lambda_j \in \mathrm{SU}(N)$, and $N>2$, the $\mathrm{Tr}(\lambda_j\rho_0)$ is not always 0. When the generators are obtained as the $\sigma_x$, one can find that $\mathrm{Tr}(\lambda_j\rho_0)\neq 0$. Example, with $\mathrm{SU}(4)$ there is the 15 generators, the generators $\lambda_0, \cdots \lambda_{5}$ satisfy $\mathrm{Tr}(\lambda_j\rho_0) = \pm 0.5$, and the generators $\lambda_0, \cdots \lambda_{5}$ is tensored by the generator $\sigma_x$ of the $\mathrm{SU}(2)$.  $\rho_2, \rho_3, \rho_4$ are the density matrices of the two-qubit, three-qubit, four-qubit complete graph states, respectively. The operators $\lambda_i$ are shown in the last section in the Appendix~\ref{sec:gen}.
\label{Fig_Gen_ave}}
\end{figure}

Next, we proof the $\mathrm{Tr}(\mathcal{H}_j \rho_0)=0$ for $\mathrm{SU}(2)$. To proof $\mathrm{Tr}(\mathcal{H}_j \rho_0)=0$, we only need to proof $\mathrm{Tr}(J_x \rho_0)=0$,  $\mathrm{Tr}(J_y \rho_0)=0$, and  $\mathrm{Tr}(J_z \rho_0)=0$, where $J_{x, y, z}= \frac{1}{2}\sum_{j=1}^n\sigma_j^{x, y, z}$ is the collective operators.
The density matrix $\rho_0$ has the expression ( \ref{eq:gra_stabi}). So we can only proof the $\mathrm{Tr}(\sigma_j^{x, y, z} \rho_0)=0$.

Firstly, for the term $\sigma^x_j \rho_0$, the $j$th position of $\sigma^x_j g_j$ is $\openone$.
However, this position of any other $g_i$ is $\openone$ or $\sigma^z$, which means
this position of $\sum_{i<j<k<\cdots <l}g_i g_j g_k\cdots g_l$ is $\sigma^x$ or $\sigma^y$,
a traceless matrix. Furthermore, since the trace of a tensor is zero if any position in
this tensor is traceless, one can find that $\mathrm{Tr}(\sigma^x_j \rho_0)=0$ for
any $j$.

Secondly, for the term $\sigma^y_j \rho_0$,  the $j$th position of $\sigma^y_j g_j$ is $\sigma^z_j$. However, this position of any other $g_i$ is $\openone$ or $\sigma^z$, which means
this position of $\sum_{i<j<k<\cdots <l}g_i g_j g_k\cdots g_l$ is $\sigma^x$ or $\sigma^y, \sigma^z$,
a traceless matrix. Furthermore, since the trace of a tensor is zero if any position in
this tensor is traceless, one can find that $\mathrm{Tr}(\sigma^y_j \rho_0)=0$ for
any $j$.

Finally, for the $\sigma^{z}_j \rho_0$, the $j$th position of the $\sigma^{z}_j g_j$ is $\sigma_j^{y}$. However, this position of any other $g_i$
is $\openone$ or $\sigma^z$, whcih means this position of $\sum_{i<j<k<\cdots <l}g_i g_j g_k\cdots g_l$ is $\sigma^{x}$ or $\sigma^y, \sigma^z$, the $\sigma^{x, y, z}$ are the traceless matrices.
Furthermore, since the trace of a tensor is zero if any position in
this tensor is traceless, one can find that $\mathrm{Tr}(\sigma^{z}_j \rho_0)=0$ for
any $j$. One can find that $\mathrm{Tr}(\mathcal{H}_j \rho_0)=0$, in which the Hermitian operator satisfies $H_k \in \mathrm{SU}(2)$.

However, when the operators $H_k \in \mathrm{SU}(N), N\geq 4$, then the $\mathrm{Tr}(\mathcal{H}_j \rho_0)$ is not always 0. Examples, when operators $H_k$ is the generators of the $\mathrm{SU}(N), N \geq 4$. The global Hamiltonian can be rewritten as $H({\bm{\theta}})=\sum_{j=1}^n \theta_j H_j$, i.e., two-qubit graph state corresponds to $\mathrm{SU}(4)$, three-qubit graph state corresponds to $\mathrm{SU}(8)$, four-qubit graph state corresponds to $\mathrm{SU}(16)$. The $\mathrm{Tr}(\rho_0 H_k)$ is not always zero   as shown in Fig.~\ref{Fig_Gen_ave}.

\section{Construct graphs with a diagonal QFIM with SU(2)}\label{Sec:digQFIM}
%============================ Figure ==============
\begin{figure}[tp]
\centering\includegraphics[width=8cm]{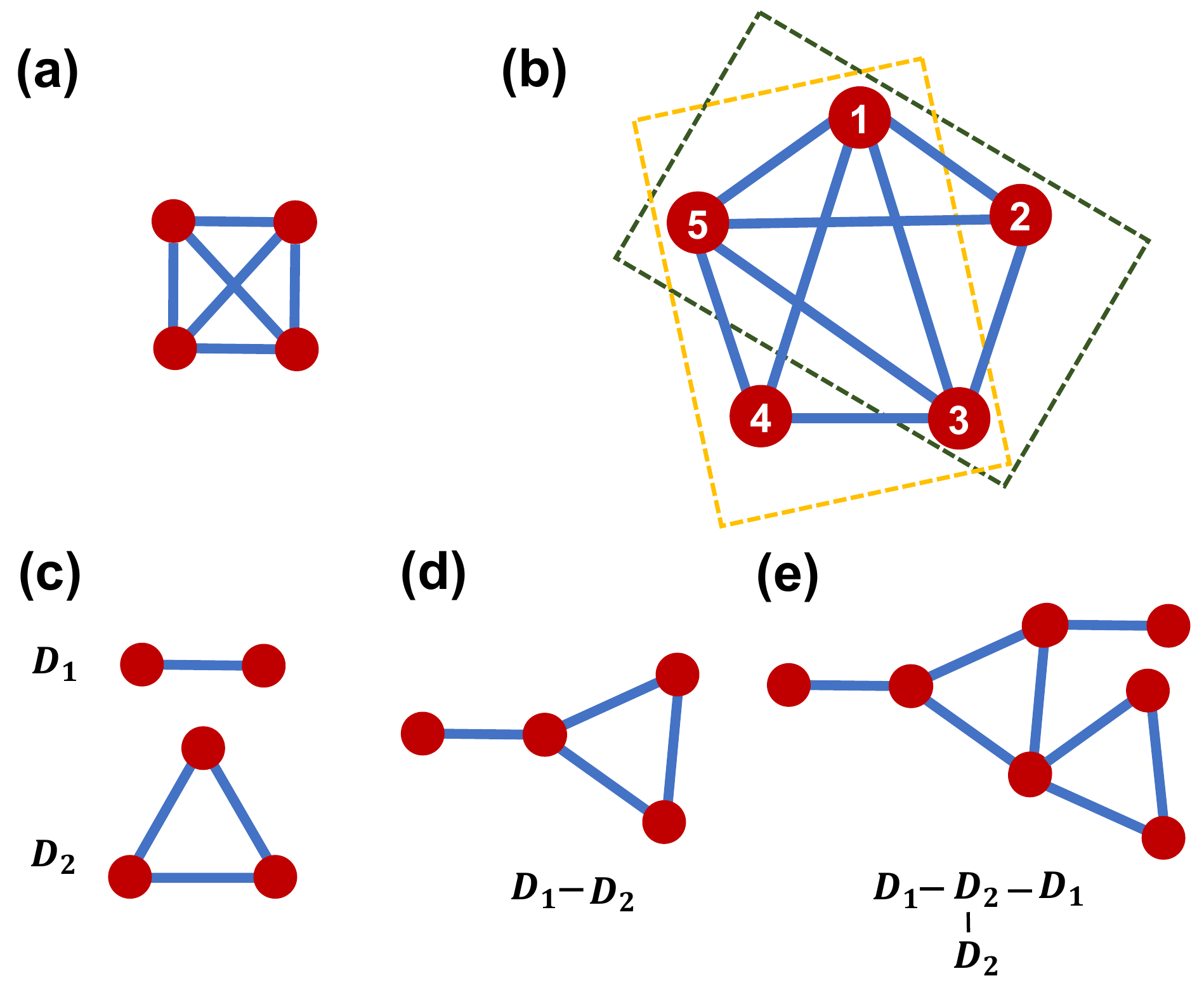}
\caption{(a) A topological number $T$ of a graph state. (b) Example with topology number $T=2$.  (c) The basic cells via the \emph{single-joint connecting rule} (SJCR, denoted by "-" in
the following). (d, e) the examples of the graphs via bell call. }
\label{fig:D_Sche}
\end{figure}
%====================================================
To construct graphs with a diagonal QFIM, we first define a set of basic cells
for the stabilizers (denoted by $\mathcal{D}$). All basic cells in $\mathcal{D}$ satisfy
(i) the QFIM is diagonal and (ii) in $\mathcal{D}$, any basic cell cannot be constructed
by other basic cells via the \emph{single-joint connecting rule} (SJCR, denoted by "-" in
the following), with which the connection of any two basic cells in a graph can only
share one joint vertex. Next, we define a topological number $T$ of a graph state as
the number of cases in which any four vertices in this graph are the neighborhood of each
other, as shown in Fig.~\ref{fig:D_Sche}(a). For example, for the graph given in
Fig.~\ref{fig:D_Sche}(b), vertices 1, 2, 3, 5 (within the dashed black line) are
the neighborhood, and so does vertices 1, 3, 4, 5 (within the dashed yellow line), therefore,
the topological number $T=2$ for this graph. In the case of $T=0$, we provide the
following theorem.

\emph{Theorem 1.} Let $\{D_1,D_2\}$ with $D_1$ and $D_2$ the basic cells
defined in Fig.~\ref{fig:D_Sche}(c),  the QFIM is the identity matrix by using constructed graphs via the SJRC.

Here we use the mathematical induction to prove the Theorem.
Denote $K$ as the number of basic cells $D_1$ and $D_2$. In the case of $K=1$,
the graphs are just $D_1$ and $D_2$, and the QFIM is obviously diagonal. Recall
that the single-joint connecting rule is denoted by "-", the graphs for $K=2$
are $D_2$-$D_2$, $D_1$-$D_1$ and $D_1$-$D_2$. The corresponding QFIMs are also
diagonal according to Eq.~(\ref{eq:Sup_QFIM_n}). Now we assume that, in the case
of $K=d$, the QFIMs of all graphs constructed by $D_1$ and $D_2$ via the single-joint
connecting rule is diagonal, and in the following, we will prove the same result
can be obtained for $K=d+1$.

\begin{figure}[tp]
\centering\includegraphics[width=7cm]{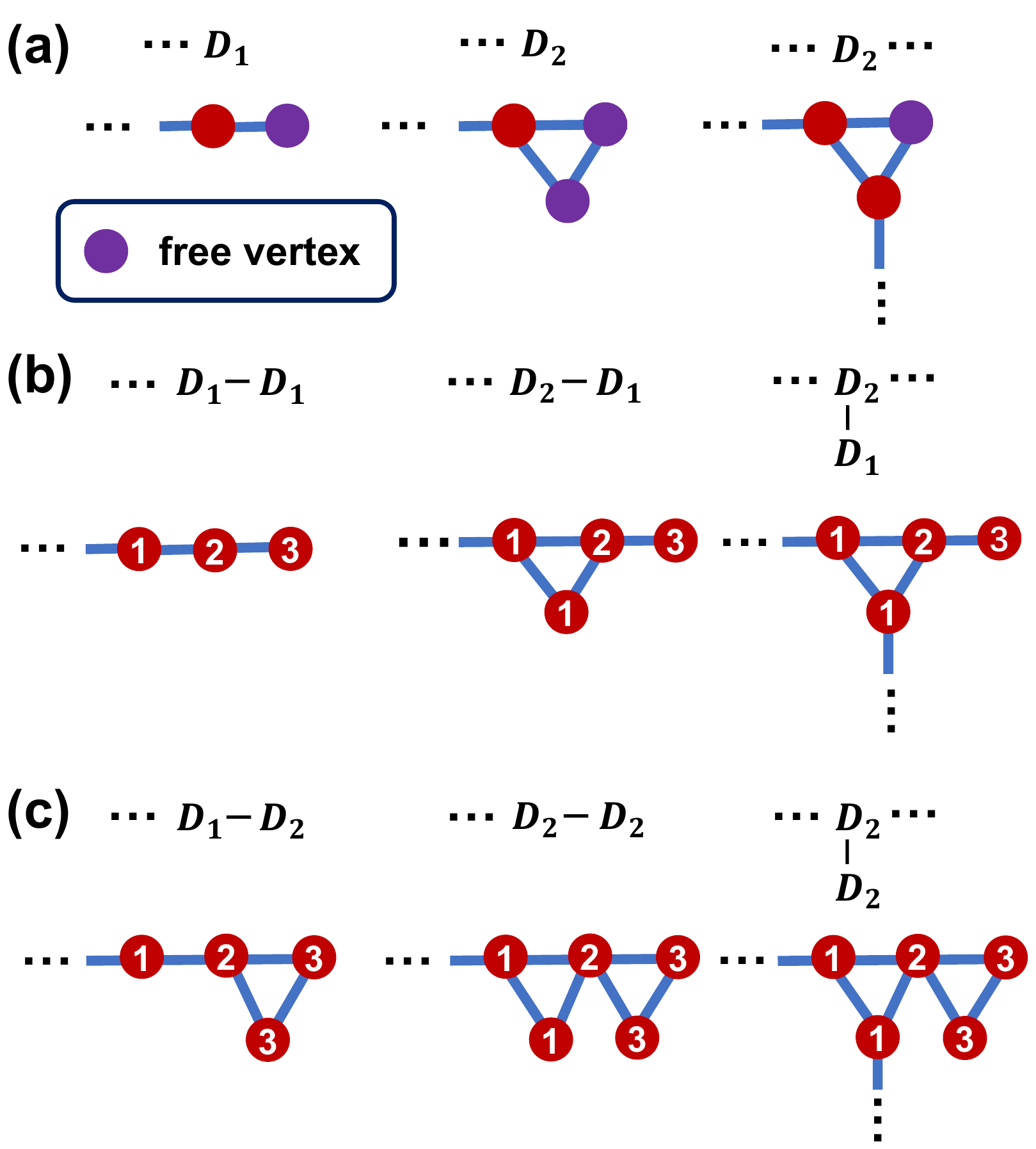}
\caption{(a) The free vertices (purple ones) in $G_d$ that are allowed by the
single-joint connecting rule to further connect a $D_1$ or $D_2$. (b) Three
scenarios of $G_d$ to connect a $D_1$. (c) Three scenarios of $G_d$ to connect
a $D_2$. Vertices 2 are the joints between $G_d$ and $D_1$ ($D_2$). Vertices 3
are new vertices in the connected new graph $G_{d+1}$, and vertices 1 are the
next nearest neighbors of vertices 3.
\label{fig:Sup_theo1}}
\end{figure}

In the case of $K=d+1$, all the graphs can be constructed by connecting graphs
with $K=d$ (denoted by $G_d$) and a $D_1$ or $D_2$. According to the single-joint
connecting rule, the vertices in $G_d$ that can perform the connections are the
free vertices (purple ones in Fig.~\ref{fig:Sup_theo1}(a)) in $D_1$ at the end
and $D_2$ in the middle or at the end. When linking $D_1$ or $D_2$ to the free vertices
in $G_d$, there exist three possible scenarios, as given in Fig.~\ref{fig:Sup_theo1}(b)
for $D_1$ and Fig.~\ref{fig:Sup_theo1}(c) for $D_2$. In all these six scenarios,
the joint vertices (labeled as 2) between $G_d$ and $D_1$ ($D_2$) have new nearest
neighbors due to the connection, however, their next nearest neighbors do not change.
Therefore, they will not contribute non-zero off-diagonal entries of the QFIM since
the QFIM for $G_d$ is diagonal. With respect to the new vertices (labeled as 3)
in the graph, the nearest neighbors of the joints (labeled as 1) in $G_d$ are
now the next nearest neighbors of vertices 3, however, they do not share
the same neighborhood since vertices 1 always connects to other vertices (including
other vertices 1) in $G_d$ besides vertices 2, yet vertices 3 only connect to
vertices 2 or other vertices 3. Hence, the connection between $G_d$ and $D_1$ ($D_2$)
via the single-joint connecting rule does not create pairs of vertices that share
the same neighborhood, and the QFIM of the new graph $G_{d+1}$ then still keeps
diagonal. The theorem is proved. $\hfill{\blacksquare}$

%============================= Figure ==========================================
%===============================================================================

\section{Calculation of the QFIM with simultaneous estimation} \label{Sec:simultaneous}

We consider the canonical case of phase estimation, where an unknown phase $\theta_{j}$ is encoded using non-interacting Hamiltonian, and the unknown parameter is $\bm{\theta}=(\theta_{1}, \theta_{2},... \theta_d )$, the unitary operator can be written as
\begin{align}
U_{\bm{\theta}}= \exp(-i \bm{H}(\bm{\theta}))
=\exp(-i \sum_{j=1}^{n} \theta_j H_j),
\end{align}
where $H_i=\frac{1}{2}\sum_j^s\lambda_{ij}$, and $\lambda_i$ is the generator of the $\mathrm{SU}(N)$. The operators $\lambda_i$ are shown in Appendix~\ref{sec:gen}. For any pure quantum state, the QFIM can reads
\begin{align}
F_{j,k}&=4\mathrm{cov}_{\rho_0}(\mathcal{H}_j,\mathcal{H}_k)
\end{align}
where $\mathcal{H}_j:= i \partial_j U^{\dagger} U= -i U^{\dagger}(\partial_j U)$, and $\mathrm{cov}(\cdot)$ is the covariance matrix between the generator $\mathcal{H}_j$ and $\mathcal{H}_k$ with parameters $\theta_j,~ \theta_k$. $\rho_0$ is the density matrix of the graph state as shown in the Eq.~(\ref{eq:gra_sta}).
And being aware of the equation
\begin{align}\label{eq:exp}
\frac{\partial e^A}{\partial x}
 &= \int_0^1 e^{(1-s)A} \frac{\partial A}{\partial x}e^{sA}\mathrm{d}s,
\end{align}
where $A$ is the Hermitian operator, and the Hermitian generator $\mathcal{H}$ can then be expresses by
\begin{align*}
\mathcal{H}_j=-\sum_{m=0} \frac{i^m}{(m+1)!}(\bm{H}(\bm{\theta})^\times)^m \partial_{\theta_j}\bm{H}(\bm{\theta}),
\end{align*}
where $H^\times(\bm{\cdot})= [H, ~\bm{\cdot}]$ is the superoperator.
For three-qubit graph state of the unitary parameterization on $\mathrm{SU}(2)$, the generators can be written as
\begin{align}
\mathcal{H}_j=& \partial_j \bm{H}(\bm{\theta})+ (1-\cos(\xi))	\frac{1}{\xi^2} [i\bm{H}(\bm{\theta}), \partial_j \bm{H}(\bm{\theta}) ]\\ \nonumber
&+ (1-\frac{\sin(\xi)}{\xi})\frac{1}{\xi^2}[i\bm{H}(\bm{\theta}),[i\bm{H}(\bm{\theta}), \partial_j \bm{H}(\bm{\theta}) ] ],
\end{align}
where $\xi= \sqrt{\theta_x^2+\theta_y^2+\theta_z^2}$ and $j =\{x, ~y, ~z\}$. For this case, the QFIM depends on the parameter values, In the limit $\theta_j \rightarrow 0$, the entries of QFIM can then
be expressed by
\begin{equation}
\mathcal{F}_{j,k}=4[\mathrm{Tr}(H_j H_k \rho_0)- \mathrm{Tr}(H_j\rho_0)\mathrm{Tr}( H_k \rho_0)].
\label{eq:QFIM_d}
\end{equation}

An example is given below to show our results.
Here we rewrite the Hamiltonian as $\bm{H}(\bm{\theta})=\theta_{1} J_1+\theta_{2}J_2+\theta_{3}J_3$ with $\theta_{1}=$ $B \sin \theta \cos \phi,  ~ \theta_{2}=B \sin \theta \sin \phi,  ~ \theta_{3}=B \cos \theta$, and the operators $J_{1, 2, 3}=\frac{1}{2}\sum_{j=1}^3\sigma_{x,y,z}^j$ are the collective operators. The QFIM is
\begin{align}
\mathcal{F}_{j,k}&=4 \langle \mathcal{H}_j\mathcal{H}_k \rangle_{\rho_0}.
\end{align}
Here, the $\mathrm{Tr}(\rho_0 \mathcal{H}_j)=0$ and $\mathrm{Tr}(\rho_0 \mathcal{H}_j\mathcal{H}_k)=\mathrm{Tr}(\rho_0 \mathcal{H}_k\mathcal{H}_j)$.
According to the Taylor expansion of trigonometric functions, the element of the QFIM with three-qubit complete graph state can be written as
\begin{align*}
  \mathcal{F}_{\theta_1, \theta_1}&=3+\frac{ 3B^{2}}{4}\left(3 \operatorname{\cos}^{2}\theta+\operatorname{\sin}\theta^{2} \operatorname{\sin}^{2}\phi\right), \\
\mathcal{F}_{\theta_2, \theta_2}&=9+\frac{3B^{2}}{4}\left(\operatorname{\cos}^{2}\theta+\operatorname{\cos}^{2}\phi \operatorname{\sin}^{2}\theta\right),\\
\mathcal{F}_{\theta_3, \theta_3}&=3+\frac{3B^{2}}{4}(2+\operatorname{\cos}(2 \phi)) \operatorname{\sin}^{2}\theta,\\
\mathcal{F}_{\theta_1, \theta_2}&=-\frac{3B}{4}\left(4 \operatorname{\cos}\theta+B \operatorname{\cos}\phi \operatorname{\sin}^{2}\theta \operatorname{\sin}\phi\right),\\
\mathcal{F}_{\theta_1, \theta_3}&= \frac{9B^{2}}{4} \operatorname{\cos}\theta \operatorname{\cos}\phi \operatorname{\sin}\theta ,\\
\mathcal{F}_{\theta_2, \theta_3}&= -\frac{3B}{4} \operatorname{\sin}\theta(-4 \operatorname{\cos}\phi+B \operatorname{\cos}\theta \operatorname{\sin}\phi).
\end{align*}
In the limit $\theta_k \rightarrow 0$, for all $k=1,2,3$, the QFIM is
\begin{equation}
\mathcal{F}=\left(\begin{array}{ccc}
4\langle J_x^2\rangle & 0 & 0 \\
0 & 4\langle J_y^2\rangle & 0 \\
0 & 0 & 4\langle J_z^2\rangle
\end{array}\right).
\end{equation}
We can find that that the precision limit  of the
three parameter estimaiton is $\mathrm{Tr}(\mathcal{F}^{-1})=\frac{2n+1}{ n^2}|_{n=3} =\frac{7}{ 9}$ with the global estimation on the $\mathrm{SU}(2)$ dynamic.

However, when the generators satisfies $[H_j, H_k]=0$, and we set the Hamiltonian is  $\bm{H}(\bm{\theta})=\frac{1}{2} (\theta_1 \sigma_1^k+\theta_2 \sigma_2^k+\theta_3 \sigma_3^k)$ for all $k= x, y, z$. In this case, we can find
the QFIM are
\begin{eqnarray*}
\mathcal{F}_x&=&\left(\begin{array}{ccc}
1 & 0 & 0 \\
0 & 1 & 0 \\
0 & 0 & 1
\end{array}\right),
\mathcal{F}_y=\left(\begin{array}{ccc}
1 & 1 & 1 \\
1 & 1 & 1 \\
1 & 1 & 1
\end{array}\right),
\mathcal{F}_z=\left(\begin{array}{ccc}
1 & 0 & 0 \\
0 & 1 & 0 \\
0 & 0 & 1
\end{array}\right).
\end{eqnarray*}
where the subscript $x, y, z$ of the QFIM indicates that the Hamiltonian operator is $\sigma^x, ~\sigma^y, ~\sigma^z$, respectively. One can find that the precision limit  of the three-parameter estimaiton is $\mathrm{Tr}(\mathcal{F}_x^{-1})=3$ with the local estimation on the $\mathrm{SU}(2)$ .

\section{Unitary multiparameter estimation: saturating the quantum Cram\'er-Rao Bound}\label{sec:Proof_QCRB}

 Now we prove the quantum Cram\'er-Rao bound could be saturated in the scenes we are considering.  For the unitary processing $U_{\bm{\theta}}$ with a pure probe graph state $\rho_0$, the necessar and sufficient condition for the attainability of quantum multiparameter Cram\'er-Rao bound is \cite{Liu2020}
\begin{align}
  \mathrm{Tr}([L_j,L_k] \rho_{\bm{\theta}})=0, ~~~\forall j, ~k ,
\end{align}
where $L_j$ is the symmetric logarithmic derivative (SLD) with respect to parameter $\theta_j$, in which is defined by $\partial_j\rho_{\bm{\theta}}=\frac{1}{2}(\rho_{\bm{\theta}}L_j + L_j \rho_{\bm{\theta}})$.
Next, we now proof the above bound equal to  \begin{equation}
  \mathrm{Tr}(\rho_0 \mathcal{H}_j\mathcal{H}_k)=\mathrm{Tr}(\rho_0 \mathcal{H}_k\mathcal{H}_j), ~~~\forall j, ~k
\end{equation}
with the initial graph state $\rho_0$. Here, the Hermitian operator is $\mathcal{H}_j:= i \partial_j U_{\bm{\theta}}^{\dagger} U_{\bm{\theta}}=-iU_{\bm{\theta}}^{\dagger}\partial_j U_{\bm{\theta}}$ with parameter $\theta_j$.
And being aware of the Eq.~(\ref{eq:exp}), the Hermitian operator
$\mathcal{H}$ can then be expressed by
\begin{align}
	\mathcal{H}_j&=-\int_{0}^1 e^{is \bm{H}(\bm{\theta})} \partial_{\theta_j}\bm{H}(\bm{\theta}) e^{-is \bm{H}(\bm{\theta})} ds,
\end{align}
and one can find
\begin{align}
U\mathcal{H}_jU^{\dagger}
	&=-\int_{0}^{1}e^{-isH}\partial_{j}H\left(\theta\right)e^{isH}ds.
\end{align}
One can find $\mathcal{R}_j \equiv -U\mathcal{H}_jU^{\dagger}$, we find
\begin{align}
	\partial_j \rho_{\bm{\theta}}=
	-i \left[ \mathcal{R}_j , \rho_{\bm{\theta}}\right],
\end{align}
where \begin{align}
	\mathcal{R}_j=\int_{0}^1 e^{-is \bm{H}(\bm{\theta})} \partial_{\theta_j}\bm{H}(\bm{\theta}) e^{is \bm{H}(\bm{\theta})} ds.
\end{align}
The SLD formula is obtained from the fact $\rho_{\bm{\theta}}^{2}=\rho_{\bm{\theta}}$ for a pure state, then $\partial_{j} \rho_{\bm{\theta}}=\rho_{\bm{\theta}} \partial_{j} \rho_{\bm{\theta}}+\left(\partial_{j} \rho\right) \rho_{\bm{\theta}}$. Compared this equation to the definition equation, it can be seen that $L_j=2 \partial_{j} \rho_{\bm{\theta}} = -2i\left[ \mathcal{R}_j , \rho_{\bm{\theta}}\right]$.
\begin{align} \nonumber
  &\mathrm{Tr}([L_j, L_k] \rho_{\bm{\theta}})\\ \nonumber
  &=\mathrm{Tr}(\rho_{\bm{\theta}}[-2i [\mathcal{R}_j, \rho_{\bm{\theta}}], -2i[\mathcal{R}_k, \rho_{\bm{\theta}}]])\\
  &=-4 \mathrm{Tr} (\rho_{\bm{\theta}} \mathcal{K}_{jk}\rho_{\bm{\theta}}),
\end{align}
where
\begin{align*}
  \mathcal{K}_{jk}\rho_{\bm{\theta}}=
  &[\mathcal{R}_j \rho_{\bm{\theta}}, \mathcal{R}_k \rho_{\bm{\theta}}] + [\rho_{\bm{\theta}} \mathcal{R}_j,  \rho_{\bm{\theta}} \mathcal{R}_k] \\
  &-[\rho_{\bm{\theta}} \mathcal{R}_j, \mathcal{R}_k\rho_{\bm{\theta}}]-[\mathcal{R}_j\rho_{\bm{\theta}}, \mathcal{R}_k\rho_{\bm{\theta}}].
\end{align*}
which implies a sufficient but not necessary condition
\begin{align}
  \mathrm{Tr}(\rho_{\bm{\theta}}[\mathcal{R}_j, \mathcal{R}_k ])=0.
\end{align}
We can find
\begin{align}
	L_{\text{eff}}=U^{\dagger}LU=2i[\mathcal{H}, \rho_0]
\end{align}
We prove the equivalence of the two conditions based on SLD and the Hermitian operator $\mathcal{H}$. Next, We proof that
\begin{equation}
  \mathrm{Im}[\mathrm{Tr}(\rho_0 \mathcal{H}_j \mathcal{H}_k)]=0,
\end{equation}
base on the equation $U \mathcal{H}_j U^\dagger = -\mathcal{R}_j$, that is to prove
\begin{equation}
  \mathrm{Im}[\mathrm{Tr}(\rho_0 \mathcal{R}_j \mathcal{R}_k)]=0,
\end{equation}
The quantum graph state is the pure state after unitary evolution $U_{\bm{\theta}}$, which can be written as  $\rho_{\bm{\theta}}= |\psi (\bm{\theta}) \rangle \langle \psi (\bm{\theta})|$. We can reconstruct QFIM by the $|\psi (\bm{\theta}) \rangle$ and $ |\partial_j \psi (\bm{\theta}) \rangle$ with parameter $\theta_j$ for all $j$. In general, the states $ |\partial_j \psi (\bm{\theta}) \rangle$ are not orthogonal to the quantum state $|\psi (\bm{\theta}) \rangle$. We can construct the matrix
\begin{eqnarray}
  G_{jk}&=&\langle \partial_j \psi (\bm{\theta}) | \partial_k \psi (\bm{\theta}) \rangle -\langle \partial_j \psi (\bm{\theta}) |  \psi (\bm{\theta}) \rangle \langle \psi (\bm{\theta}) | \partial_k \psi (\bm{\theta}) \rangle\\ \nonumber
  &=&\mathrm{Tr} (\rho_0 U_{\bm{\theta}}^\dagger  \mathcal{H}_j \mathcal{H}_k U_{\bm{\theta}} )-\mathrm{Tr} (\rho_0 U_{\bm{\theta}}^\dagger  \mathcal{H}_j  U_{\bm{\theta}} ) \mathrm{Tr} (\rho_0 U_{\bm{\theta}}^\dagger  \mathcal{H}_k U_{\bm{\theta}} )\\
  &=&\mathrm{Tr}(\rho_0 \mathcal{R}_j \mathcal{R}_k)- \mathrm{Tr}(\rho_0 \mathcal{R}_j)\mathrm{Tr}(\rho_0\mathcal{R}_k),
\end{eqnarray}
where $|\partial_j \psi (\bm{\theta}) \rangle = -i \mathcal{H}_j|\psi (\bm{\theta}) \rangle $. It is noticed that, according to Eq.~(\ref{eq:QFIM_d}), the QFIM is the real and symmetry matrix. The operator $\mathcal{R}$ is the Hermitian operator, that $\mathrm {Tr}(\rho_0 \mathcal{R})$ is the real number, we can find $\mathrm{Tr}(\rho_0 \mathcal{R}_j \mathcal{R}_k)$ is always the real number.

\section{The generators of the SU(N)}\label{sec:gen}
The $N^2-1$ traceless non-diagonal and diagonal symmetric and antisymmetric $\mathrm{SU}(N)$ generators are obtained as the following equations~\cite{Omolo2018}.\\
$N=2$:
\begin{equation*}
\begin{aligned}
\sigma_x &=\left(\begin{array}{cccc}
0 & 1   \\
1 & 0  \\
\end{array}\right),  \sigma_y=\left(\begin{array}{cccc}
0 & -i \\
i & 0  \\
\end{array}\right),
\sigma_z &=\left(\begin{array}{cccc}
1 & 0 \\
0 & -1 \\
\end{array}\right).
\end{aligned}
\end{equation*}
$N=4$:
\begin{equation*}
\begin{aligned}
&\lambda_{0} =\left(\begin{array}{cccc}
0 & 1 & 0 & 0 \\
1 & 0 & 0 & 0 \\
0 & 0 & 0 & 0 \\
0 & 0 & 0 & 0
\end{array}\right),  \lambda_{1}=\left(\begin{array}{cccc}
0 & 0 & 1 & 0 \\
0 & 0 & 0 & 0 \\
1 & 0 & 0 & 0 \\
0 & 0 & 0 & 0
\end{array}\right),\\
&\lambda_{2} =\left(\begin{array}{llll}
0 & 0 & 0 & 1 \\
0 & 0 & 0 & 0 \\
0 & 0 & 0 & 0 \\
1 & 0 & 0 & 0
\end{array}\right),
\lambda_{3}=\left(\begin{array}{cccc}
0 & 0 & 0 & 0 \\
0 & 0 & 1 & 0 \\
0 & 1 & 0 & 0 \\
0 & 0 & 0 & 0
\end{array}\right),\\
&\lambda_{4} =\left(\begin{array}{cccc}
0 & 0 & 0 & 0 \\
0 & 0 & 0 & 1 \\
0 & 0 & 0 & 0 \\
0 & 1 & 0 & 0
\end{array}\right),
\lambda_{5}=\left(\begin{array}{cccc}
0 & 0 & 0 & 0 \\
0 & 0 & 0 & 0 \\
0 & 0 & 0 & 1 \\
0 & 0 & 1 & 0
\end{array}\right),
\end{aligned}
\end{equation*}
and
\begin{equation*}
\begin{aligned}
&\lambda_{6}=\left(\begin{array}{cccc}
  0 & -i & 0 & 0 \\
  i & 0 & 0 & 0 \\
  0 & 0 & 0 & 0 \\
  0 & 0 & 0 & 0
\end{array}\right),
\lambda_{7} =\left(\begin{array}{cccc}
0 & 0 & -i & 0 \\
0 & 0 & 0 & 0 \\
i & 0 & 0 & 0 \\
0 & 0 & 0 & 0
\end{array}\right),\\
 &\lambda_{8}=\left(\begin{array}{cccc}
0 & 0 & 0 & -i \\
0 & 0 & 0 & 0 \\
0 & 0 & 0 & 0 \\
i & 0 & 0 & 0
\end{array}\right),
\lambda_{9} =\left(\begin{array}{cccc}
0 & 0 & 0 & 0 \\
0 & 0 & -i & 0 \\
0 & i & 0 & 0 \\
0 & 0 & 0 & 0
\end{array}\right), \\
&\lambda_{10}=\left(\begin{array}{cccc}
0 & 0 & 0 & 0 \\
0 & 0 & 0 & -i \\
0 & 0 & 0 & 0 \\
0 & i & 0 & 0
\end{array}\right),
\lambda_{11}=\left(\begin{array}{cccc}
0 & 0 & 0 & 0 \\
0 & 0 & 0 & 0 \\
0 & 0 & 0 & -i \\
0 & 0 & i & 0
\end{array}\right),
\end{aligned}
\end{equation*}
and
\begin{equation*}
\begin{aligned}
&\lambda_{12} =\left(\begin{array}{cccc}
  1 & 0 & 0 & 0 \\
  0 & -1 & 0 & 0 \\
  0 & 0 & 0 & 0 \\
  0 & 0 & 0 & 0
  \end{array}\right),
  \lambda_{13}=\frac{1}{\sqrt{3}}\left(\begin{array}{cccc}
 1 & 0 & 0 & 0 \\
 0 & 1 & 0 & 0 \\
 0 & 0 & -2 & 0 \\
 0 & 0 & 0 & 0
 \end{array}\right).
 \end{aligned}
 \end{equation*}
 and
 \begin{equation*}
 \begin{aligned}
 &\lambda_{14}=\frac{1}{\sqrt{6}}\left(\begin{array}{cccc}
1 & 0 & 0 & 0 \\
0 & 1 & 0 & 0 \\
0 & 0 & 1 & 0 \\
0 & 0 & 0 & -3
\end{array}\right).
\end{aligned}
\end{equation*}
According to the methods shown in the literature~\cite{Omolo2018}, we can write high-dimensional generators of the $\mathrm{SU}(2^n)$.

\end{document}